\documentclass[runningheads]{comsis2}

\pdfoutput=1
\def\journalissue{Computer Science and Information Systems 00(0):0000--0000}
\def\paperidnum{https://doi.org/10.2298/CSIS123456789X}
\PassOptionsToPackage{hyphens}{url}\usepackage{hyperref}
\usepackage{multirow}
\usepackage{booktabs}
\usepackage{orcidlink}
\usepackage{graphicx}
\usepackage{pgfplots,pgfplotstable}
\usetikzlibrary{patterns,patterns.meta}
\usepackage[normalem]{ulem}
\usepackage{paralist}
\usepackage{subcaption}
\usepackage{stix}  
\usepackage{ifsym}
\usepackage[colorinlistoftodos]{todonotes}
\usepackage{enumitem}
\usepackage{comment}
\usepackage{cite}
\usepackage{amsmath}  

\usepackage{listings}
\usepackage{xcolor}
\definecolor{codegreen}{rgb}{0,0.6,0}
\definecolor{codegray}{rgb}{0.5,0.5,0.5}
\definecolor{codepurple}{rgb}{0.58,0,0.82}
\definecolor{backcolour}{rgb}{0.95,0.95,0.92}

\lstdefinestyle{gql}{
    backgroundcolor=\color{backcolour},   
    commentstyle=\color{codegreen},
    keywordstyle=\color{magenta},
    numberstyle=\tiny\color{codegray},
    stringstyle=\color{codepurple},
    basicstyle=\ttfamily\scriptsize,
    breakatwhitespace=false,         
    breaklines=true,                 
    captionpos=b,                    
    keepspaces=true,                 
    numbersep=5pt,                  
    showspaces=false,                
    showstringspaces=false,
    showtabs=false,                  
    tabsize=2
}

\newtheorem{atheorem}{Theorem}

\title{Siren Federate: Bridging Document, Relational, and Graph Models for Exploratory Graph Analysis%
\footnote{This manuscript is an extended version of the paper ``Siren Federate: Bridging the Gap between Document and Relational Data Systems for Efficient Exploratory Graph Analysis''~\cite{catena2024siren} which has appeared in the proceedings of the 28th International Symposium on Database Engineered Applications (IDEAS 2024). This is the pre-print version submitted for review to the Computer Science and Information Systems (ComSIS) journal.}}

\titlerunning{Siren Federate}

\author{
	Georgeta Bordea\orcidlink{0000-0001-9921-8234}\inst{1} \and
	St\'ephane Campinas\orcidlink{0009-0008-3058-5376}\inst{2} \and
	Matteo Catena\orcidlink{0000-0002-5571-8269}\inst{2} \and
	Renaud Delbru\orcidlink{0009-0007-7759-693X}\inst{2}
}

\institute{
	L3i, La Rochelle University -- La Rochelle, France\\
	\email{name.surname@univ-lr.fr}\\
	\url{https://l3i.univ-larochelle.fr/}
	\and
	Siren -- Galway, Ireland\\
	\email{name.surname@siren.io}\\
	\url{https://siren.io/}
}

\begin{document}

\maketitle

\begin{abstract}
Investigative workflows require interactive exploratory analysis on large heterogeneous knowledge graphs. Current databases show limitations in enabling such task. This paper discusses the architecture of Siren Federate, a system that efficiently supports exploratory graph analysis by bridging document-oriented,  relational and graph models. Technical contributions include distributed join algorithms, adaptive query planning, query plan folding, semantic caching, and semi-join decomposition for path query. Semi-join decomposition addresses the exponential growth of intermediate results in path-based queries. Experiments show that Siren Federate exhibits low latency and scales well with the amount of data, the number of users, and the number of computing nodes.

\vspace{6pt}\textbf{Keywords:} Exploratory Graph Analysis, Knowledge Graph, Database and Information System Architecture, Distributed Join Algorithms, Document-oriented Database.
\end{abstract}

\section{Introduction}

Investigative Intelligence encompasses domains such as law enforcement, financial compliance, cyber-threat analysis, and investigative journalism, in which professionals uncover hidden patterns and relationships by analyzing diverse and interconnected data sources~\cite{BBS17,AOH09,kejriwal2018investigative,balalau2024graph,jamil2025online}. This analysis enables the detection of emerging threats, verification of claims, and identification of concealed connections, contributing to transparency, accountability, and informed decision-making in contexts involving corruption, misinformation, and complex criminal activities. These scenarios require integrating structured records, semi-structured logs, unstructured text, and increasingly, multi-modal content such as images and videos. A central challenge lies in enabling rapid, flexible, and iterative analysis over massive volumes of heterogeneous data.

Knowledge Graphs (KGs)~\cite{hogan2021knowledge} provide a unifying abstraction for integrating such diverse data into a single graph representation, where vertices represent entities and edges their relationships. This unified view enables analysts to ``connect the dots'' across modalities and data silos. However, scaling exploratory graph analysis to billions of entities -- while maintaining interactive responsiveness -- remains an open challenge.

Current database technologies present significant limitations for investigative workflows. Native graph databases excel at localized traversals but struggle with global analytical workloads and scalability. Relational databases provide robust structured data operations but lack flexibility for heterogeneous data and graph processing. Both approaches face a common challenge with path enumeration, where intermediate result explosion undermines performance at scale.

To address these challenges, we present \emph{Siren Federate}, a system that integrates relational and graph querying capabilities into Elasticsearch, a distributed Information Retrieval (IR) system~\cite{GT2015}. By leveraging Elasticsearch's advanced text search, multi-dimen\-sional search, vector search, and horizontal scalability, Siren Federate extends these strengths by incorporating distributed join algorithms, adaptive query planning, and semantic caching strategies. The result is a system that supports iterative exploration of large, multi-modal knowledge graphs, providing the multi-hop expansions and relational filtering essential for investigative workflows, all within interactive, sub-second to second response times.

This manuscript describes how Siren Federate bridges document-oriented, relational and graph models. The architecture presented here has been refined through a decade of R\&D and production deployments, including installations that span hundreds of nodes and process petabyte-scale datasets. Rather than describing a theoretical system or research prototype, we share insights and optimizations derived from addressing real-world investigative intelligence challenges at scale. 

We first outline the investigative intelligence domain and practical requirements in Sec.~\ref{sec:investigative}. We then present an overview of related work in Sec.~\ref{sec:related-work}, which examines general limitations of current database technologies and specific approaches to path query processing. Next, Sec.~\ref{sec:bridge} discusses how to bridge the document, relational, and graph models to enable exploratory graph analysis, while Sec.~\ref{sec:arch} presents Siren Federate’s architecture, focusing on its distributed join algorithms, adaptive query optimization, query plan folding, and semantic caching. In Sec.~\ref{sec:sjd}, we introduce a novel \emph{Semi-Join Decomposition} technique that addresses the exponential growth of intermediate results in path-based queries. Finally, in Sec.~\ref{sec:evaluation}, we evaluate Siren Federate's performance using large synthetic datasets and the LDBC Financial Benchmark. These experiments demonstrate our system's capability to support complex investigative scenarios while maintaining responsiveness suitable for real-time exploration. By unifying IR, relational, and graph querying, Siren Federate offers a foundation on which investigative intelligence systems can efficiently handle increasingly complex and diverse data.

\section{Investigative Intelligence Background}
\label{sec:investigative}

This section provides essential background on investigative intelligence workflows and their requirements for database systems. We first outline the domain and its distinctive data exploration patterns, then identify the specific technical capabilities required to support these workflows.

\subsection{Domain Overview}
\label{sec:investigative:overview}

Investigative workflows in law enforcement and cybersecurity increasingly rely on large knowledge graphs to detect threats by connecting data involving suspects, organizations, financial transactions, and network events. Similar requirements arise in investigative journalism and financial compliance. Journalists must collate and verify evidence from data leaks, government archives, or user-generated content; compliance teams need to trace suspicious transaction patterns across multiple accounts to flag potential money laundering or fraud. Malicious actors exploit the increasing volume and complexity of data to blend in and operate undetected. Therefore, analysts require the capability to iteratively explore complex chains of relationships hidden within massive volumes of heterogeneous data, traversing across textual, numeric, relational, and multimedia modalities.

Knowledge Graphs (KGs)~\cite{hogan2021knowledge} serve as the foundation for these workflows by integrating diverse data sources --- structured data (e.g., relational tables), semi-structured data (e.g., JSON, XML), and unstructured data (e.g., text, images, multimedia) --- into a unified graph representation. This consolidated view enables analysts to ``connect the dots'' across data types, facilitating the identification of threats, cross-border criminal activity, or hidden corruption within complex corporate structures. KGs also enhance transparency by visualizing relationships, providing reasoning pathways, and ensuring traceable provenance -- critical components in building and validating investigative narratives.

Investigations rarely proceed as single, linear queries. Instead, they involve exploratory, iterative analysis aimed at discovering hidden patterns and generating new leads~\cite{Lis+22}. Analysts typically begin with limited information -- such as a suspicious keyword in contractual documents or an unusual shipping reference -- and progressively expand their exploration across structured data, full-text documents, and multi-modal features to uncover additional evidence and refine their hypotheses. The investigative system must guide users through iterative searching, filtering, and data drilling, enabling rapid identification of entities of interest. Graph analytics capabilities, such as path-finding, centrality, and community detection, further assist in discovering relevant subgraphs and patterns. 

Siren's platform~\cite{conf/iir/CampinasCD23} addresses these investigative needs by combining multiple data interaction paradigms -- including search, analytic dashboards, set-to-set navigation, and graph visualization -- into a unified exploration model. For example, in a Signals Intelligence scenario, individuals, cellphone data, call records, text messages, and network cells form a complex, interconnected graph. Set-to-set navigation~\cite{oren2006extending} (a type of relational faceted navigation) enables analysts to easily navigate among these interrelated datasets, as applying filters to one set dynamically updates all related entities, supporting iterative exploration. An analyst may start with a textual clue (e.g., a suspicious message), then iteratively expand towards linked phone records, geolocations, or associated media. Graph visualization further helps in understanding the interconnected datasets, identifying clusters, and answering targeted questions such as \emph{``Which individuals own these phones? Are they connected to the same network cells? Are they part of a coordinated group?''}.

\subsection{Requirements for Investigative Intelligence Systems}
\label{sec:investigative:requirements}

To support investigative workflows, exploratory graph analysis system must handle diverse workloads that combine multiple querying paradigms:
\begin{enumerate}
	
	\item Information Retrieval workloads, including: 
	\begin{inparaenum}[(a)]
		\item Standard and advanced full-text searches over textual documents (such as social media posts, open web data, forensic reports, etc.), including keyword-based search, fuzzy matching, wildcard queries, and phonetic similarity to account for misspellings or alternate spellings. 
		\item Relevance ranking and highlighting for prioritizing results and identifying key passages. 
		\item Semantic and vector-based searching across multi-modal data. 
		\item Multi-dimensional search, including geo-spatial (e.g., GPS coordinates, polygon-based region), temporal (e.g., event histories, timestamps), and numeric ranges (e.g., IP intervals, financial amounts). 
	\end{inparaenum}
	
	\item Relational and graph database workloads \cite{besta2023demystifying}, such as: 
	\begin{inparaenum}[(a)]
		\item OLTP-style queries for rapid, localized retrieval of nodes and edges (e.g., fetch all immediate connections of a particular entity). 
		\item OLAP-style analytical queries that aggregate data over large graph segments, such as frequency distributions or statistical summarizations over extensive graph regions. 
		\item Neighborhood queries and localized traversals within limited hops (degrees of separation). 
		\item Path-finding queries and traversals across long graph paths (e.g., shortest paths). 
		\item Global graph analytical operations such as pattern matching, keyword-based subgraph searching, community detection, centrality analyses. 
	\end{inparaenum}
	
	\item Hybrid exploratory workloads, involving combinations of IR and relational-graph queries, such as: 
	\begin{inparaenum}[(a)] 
		\item Multi-hop filtered expansions constrained by textual relevance, semantic similarity, or attribute conditions. 
		\item Interactive drill-down and faceted exploration, integrating relational joins and full-text or semantic filtering to iteratively narrow down datasets based on user-selected criteria. 
		\item Cross-modal correlation, where vector searches on one data modality dynamically filter relational expansions on another modality (such as numeric financial records or geospatial locations). 
	\end{inparaenum}
	
\end{enumerate}

These workloads require flexible data modeling and efficient query execution across heterogeneous data. Additionally, investigative systems must ensure fast response time, as slow interactions can disrupt cognitive flow~\cite{LJ14}. Supporting iterative, multi-step queries over large knowledge graphs while preserving sub-second to second responsiveness remains a critical challenge. Moreover, analysts increasingly require multi-modal querying across embeddings derived from text, images, or audio. Despite their importance, such capabilities are underrepresented in KG systems and remain an active research area~\cite{khan2023knowledge}.

Given these requirements, it is important to examine how current database technologies address these challenges and where they fall short. In the next section, we review related work covering both general limitations of existing database technologies and specific approaches to path query processing, which will provide context for our approach.

\section{Related Work}
\label{sec:related-work}

This section reviews existing database technologies and query processing approaches relevant to the challenges of exploratory graph analysis in investigative intelligence. We first examine general limitations of current database technologies for handling heterogeneous knowledge graphs at scale, then provide a focused analysis of specific techniques for path query evaluation.

\subsection{Limitations of Current Database Technologies}
\label{sec:related-work:current-db}

Despite advances, current graph and relational database solutions often fall short in meeting the requirements of investigative workflows~\cite{besta2023demystifying}. 

Native graph databases excel at localized traversals but may not be effective for global analytical workloads. \cite{pokorny2015graph,traris2021path} noted challenges in scaling to very large graphs and handling complex pattern-matching queries. \cite{manolescu2023full} highlight limitations in current graph query languages and call for richer query paradigms combining structured and IR-style search. Many graph databases provide limited built-in support for global graph analytics, typically offering only basic functions while requiring external libraries for advanced algorithms~\cite{traris2021path}. They also often lack robust semantic similarity search capabilities~\cite{vrgoc2024millenniumdb}, although systems like Neo4j have recently added vector indexing.

Relational databases handle structured data and global analytics effectively, but they lack flexibility in managing heterogeneous data and exhibit limited native support for graph and advanced text processing~\cite{nosqlsurvey}. When used for graph workloads, they produce ``huge redundant intermediate data'' during join operations~\cite{shen2023bridging}, with multi-hop queries degrading by ``several orders of magnitude'' compared to native graph approaches. 

Multi-model databases attempt to integrate multiple data models (key-value, document, graph), but originating from specific data paradigms frequently leads to inefficiencies in complex workloads~\cite{zhang2019unibench}. Cross-model query optimization presents significant challenges, particularly for hybrid queries spanning data models. Achieving optimal performance across relational, graph, and full-text search remains difficult due to challenges in query processing, schema design, and indexing~\cite{multimodalDBs}. Finally, polyglot architectures -- combining specialized systems like text search engines, graph databases, and vector stores -- add operational complexity, fragment user workflows, and create challenges in maintaining data consistency and coordinated queries~\cite{NBM17, kiehn2022polyglot}.

As a result, analysts face operational complexity, fragmented user experiences, and missed investigative insights when queries cannot seamlessly span data modalities. To the best of our knowledge, no single system fully integrates advanced IR over multi-modal data with scalable relational joins and graph analytics. In particular, joins are fundamental to robust graph analysis~\cite{ten2023duckpgq}. We argue that combining the strengths of IR, relational, and graph paradigms within a unified framework is a balanced solution to these challenges.

This motivated our adoption of Elasticsearch, which offers a flexible data model, horizontal scalability, and advanced IR capabilities -- including full-text search, multidimensional indexing, and vector similarity. However, Elasticsearch lacks native support for query-time joins across shards, requiring indexing-time preprocessing and often resorting to denormalization.\footnote{\url{https://www.elastic.co/guide/en/elasticsearch/reference/8.17/joining-queries.html}} These constraints only adequately supports simple hierarchical relationships, and are inadequate for complex graph data. Siren Federate addresses these challenges by integrating scalable, query-time distributed joins and graph capabilities into Elasticsearch. Sec.~\ref{sec:bridge}~and~\ref{sec:arch} detail our integration approach and system architecture that enable these capabilities.

\subsection{Path Query Techniques and Limitations}
\label{sec:related-work:path-query}

Path queries represent a specific challenge within the broader limitations of current database technologies. Existing approaches to path enumeration fall into two main categories: one rooted in graph traversal, through variants of the breadth-first search (BFS), and another grounded in relational algebra, using join operators to express multi-hop relationships.

\subsubsection{Graph Traversal (BFS-Based) Approaches}

BFS and its derivatives are the most intuitive methods for computing shortest paths~\cite{moore1959shortest}. These algorithms proceed in iterative ``waves'', expanding the frontier at each step. To enumerate all shortest paths between a source and target node, BFS is often combined with a backtracking phase over a shortest-path direct acyclic graph (DAG) built during traversal~\cite{vrgovc2022evaluating}. Following the graph model defined in Sec.~\ref{sec:bridge}, a typical BFS-based workflow involves:
\begin{inparaenum}[(1)]
	\item Starting from a source node $u$, retrieving immediate neighbors based on a field-matching condition;
	\item Appending these neighbors into a frontier queue while maintaining a visited set;
	\item Iteratively expanding each node in the frontier, maintaining parent pointers to enable backtracking for path reconstruction;
	\item Continuing until the target node $v$ is reached, or the maximum path length $L$ is reached. 
\end{inparaenum}
Even after finding the target vertex, the algorithm must continue exploring any remaining frontier nodes at the current depth to ensure all shortest paths of equal minimal length are identified. As traversal proceeds in a dense graph, each level may yield an exponential number of partial paths to track. These intermediate structures must be maintained until the algorithm determines which paths are truly minimal. Even with compact representations like predecessor DAGs, the memory footprint of intermediate state grows rapidly. 

Native graph databases like Neo4j\footnote{\url{https://neo4j.com/docs/graph-data-science/current/algorithms/pathfinding/}} and Memgraph\footnote{\url{https://memgraph.com/docs/advanced-algorithms/deep-path-traversal}} implement built-in functions to enumerate all minimal-length paths between two nodes. Internally, these engines rely on BFS-style traversal, maintaining a shortest-path DAG and exploring it to produce all valid path permutations. \cite{Martens2023} highlight the critical issue of exponential intermediate path sets in graph database queries, proposing compact path multiset representations to manage intermediate result explosion. \cite{Peng2021} similarly demonstrate the combinatorial explosion of candidate paths in hop-constrained path enumeration, proposing efficient polynomial-delay algorithms to mitigate this issue. \cite{Hao2021} extend BFS approaches to billion-scale distributed graphs by combining BFS and DFS searches with aggressive pruning methods to control intermediate result sizes. \cite{Lai2021} propose an FPGA-based batching strategy using DFS principles to curb BFS-related exponential memory overhead effectively.

Although recent approaches offer meaningful improvements through indexing, pruning, and hardware optimizations, BFS-based traversal remains fundamentally limited by exponential growth in intermediate state and runtime, particularly in dense graphs with high branching factors. The Resonance Algorithm~\cite{liu2023resonance}, while innovative in its wave propagation technique, also suffers from maintaining extensive intermediate path states, offering only limited mitigation of path explosion.

\subsubsection{Relational Algebra (Join-Based) Approaches}

An alternative class of methods models path traversal as a sequence of relational joins $D_1 \bowtie \dots \bowtie D_{l+1}$ where each join captures an edge in the traversal path. Formally, given a path length $l \leq L$, the query would be structured as $D_1 \bowtie \dots \bowtie D_{l+1}$ where each $D_i$ represents an index participating in the query. Typically, the indices involved (${D_1, \dots, D_{l+1}}$) and their order are determined by analyzing the underlying ontology or schema that defines relationships between document types (see Sec.~\ref{sec:bridge}). Executing such inner-join chains explicitly enumerates all possible paths connecting the source node $u \in D_1$ and the target node $v \in D_{l+1}$. 

This approach benefits from mature query optimization infrastructure, data locality, and bulk-processing efficiencies inherent in relational algebra. For instance, \cite{jindal2015graph} explored using standard relational operations to perform path traversal. Evaluating a shortest-path query typically involves issuing one join chain per tested path length $l$, terminating when a match is found or the maximum length $L$ is reached. However, relational joins suffer from intermediate result explosion due to Cartesian product semantics. At each join step, the engine computes all combinations of matching documents between consecutive indices, resulting in exponential growth in the number of partial paths. Specifically, for a query of length $l$, intermediate results at step $i$ can reach $O(b^i)$, where $b$ is the average out-degree~\cite{modernAI}. Although many of these paths are ultimately discarded, they are nonetheless materialized and carried forward, incurring significant costs.

\cite{yakovets2016query} address this explosion through cost-based query planning specifically for recursive property paths, while \cite{Tziavelis2021Beyond} propose factorized representations to efficiently enumerate join results, significantly reducing redundancy. \cite{Mulder2025Optimizing} introduce optimization techniques to prune redundant paths early and strategically reuse intermediate results in navigational graph queries. \cite{Angles2025PathAlgebra} further establishes algebraic foundations, highlighting theoretical limitations of unrestricted recursive queries. It is important to note that while worst-case optimal join algorithms, such as those by \cite{ngo2018worst}, tightly bound intermediate result sizes in conjunctive pattern matching, they do not inherently address recursive or iterative path enumeration challenges.

Graph-oriented relational frameworks such as \cite{gao2011relational,paradies2015graphite,zhao2017all} adopt a hybrid approach by implementing graph operators within relational database systems, leveraging existing SQL engines and optimizers. Although this approach enables BFS-style graph expansions using database primitives, it does not alleviate the fundamental issue of exponential path enumeration. PathEnum~\cite{sun2021pathenum} constructs a lightweight query-time index based on vertex distances to prune invalid edges before enumeration, significantly outperforming previous methods by reducing the search space, yet still potentially materializing a substantial number of invalid partial results.

\subsubsection{Partial-Path Explosion}

Both classes of approaches face the same fundamental challenge: they materialize a large set of intermediate path candidates, only to later discard many of them after determining they do not meet final criteria. This materialization overhead becomes prohibitive in large-scale graphs, motivating our exploration of semi-join decomposition as an alternative approach. These specific limitations directly motivate our semi-join decomposition technique introduced in Sec.\ref{sec:sjd}.

\section{Bridging the Document, Relational, and Graph Models}\label{sec:bridge}

\begin{figure}
	\centering
	\begin{minipage}[t]{.5\textwidth}
		\centering
		\includegraphics[width=.95\textwidth]{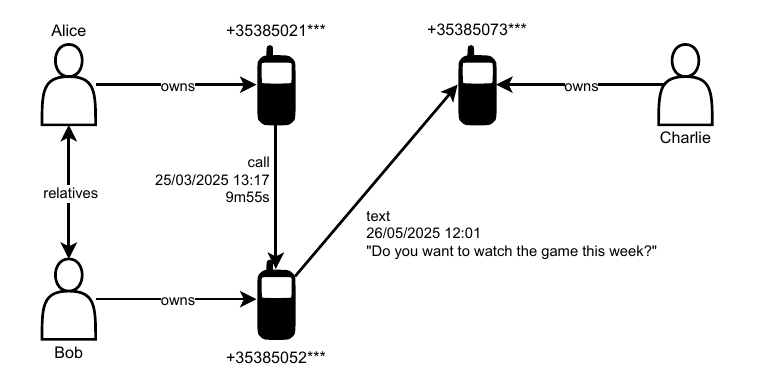}
		\caption{An example graph}
		\label{fig:graph}
	\end{minipage}%
	\begin{minipage}[t]{.5\textwidth}
		\centering
		\includegraphics[width=.95\textwidth]{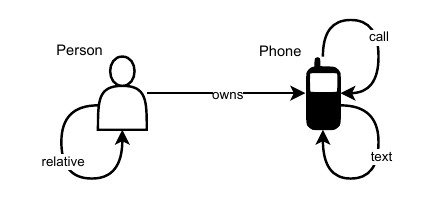}
		\caption{The related ontology} 
		\label{fig:ontology}
	\end{minipage}
\end{figure}

Knowledge graphs are powerful constructs for representing real-world entities and their relations via vertices and edges. Fig.~\ref{fig:graph} shows a graph with three people (Alice, Bob, and Charlie), where Alice and Bob are relatives. Each person owns a cellphone contract associated with a phone number. On a certain date, Alice's phone called Bob for about 10 minutes. On another occasion, Bob's phone texted Charlie about a sport match. These entities are represented as vertices in the graph, while edges depict the relationships between them (e.g., ``being relatives'', ``making a call'', etc.).

More formally, a knowledge graph can be seen as a directed graph $G(V,E)$ with $V=\{v_1,\dots,v_{|V|}\}$ being the set of nodes or vertices (representing entities), and $E$ being the set of edges (representing relations). An edge $e_{vw} \in E$ is a directed link from a source node $v$ to a target node $w$. The types of entities and relation that can occur in a knowledge graph are described by its \textit{ontology}. For example, Fig.~\ref{fig:ontology} shows that our graph can contains two types of entities -- people and phones. In terms of relations, the ontology shows that people can be relatives to each other and can own phones, which can call and text.

\begin{figure}
	\centering
	\begin{minipage}[t]{.5\textwidth}
		\centering
		\includegraphics[width=.95\textwidth]{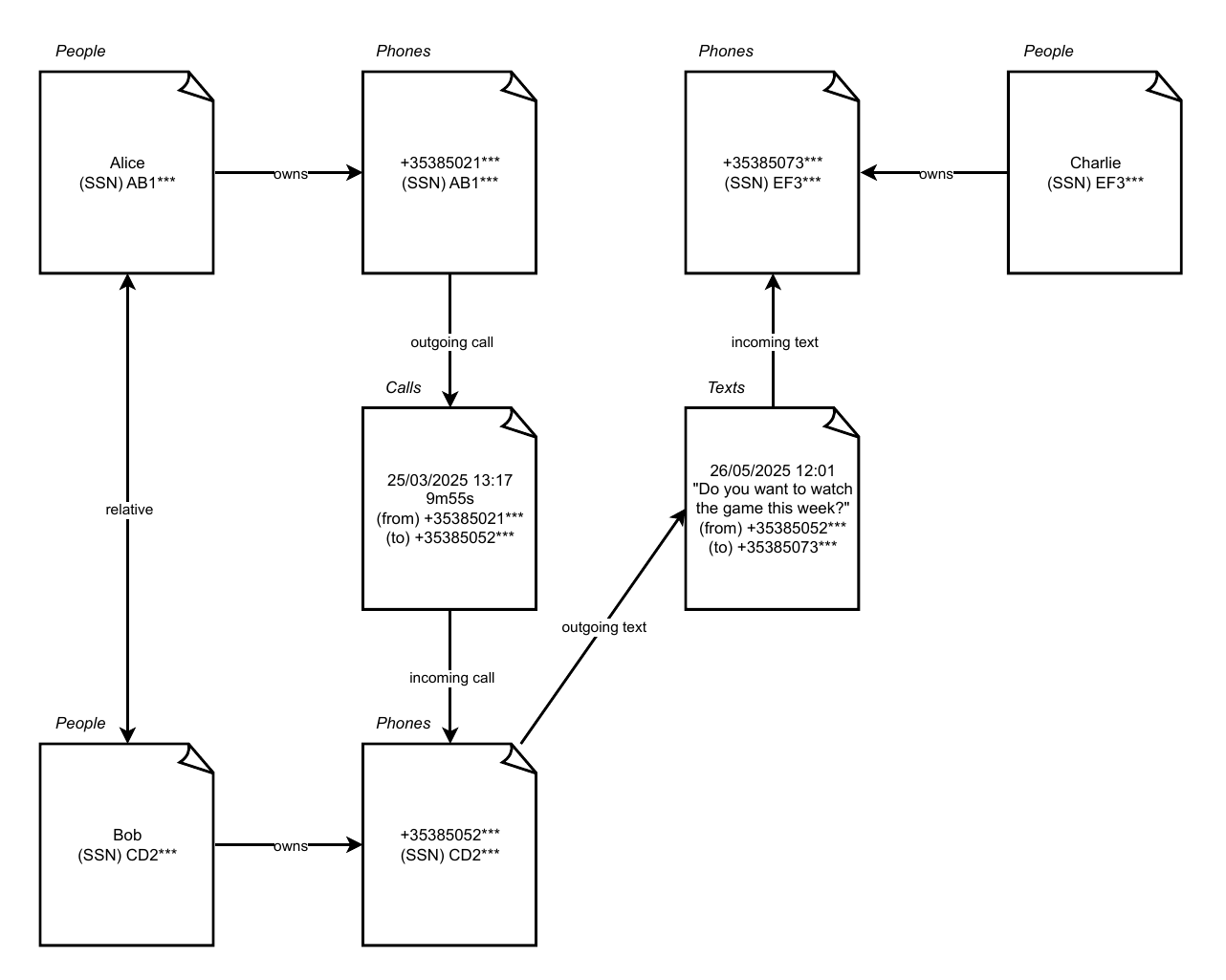}
		\caption{The example graph in a document-oriented store}
		\label{fig:graph-as-docs}
	\end{minipage}%
	\begin{minipage}[t]{.5\textwidth}
		\centering
		\includegraphics[width=.95\textwidth]{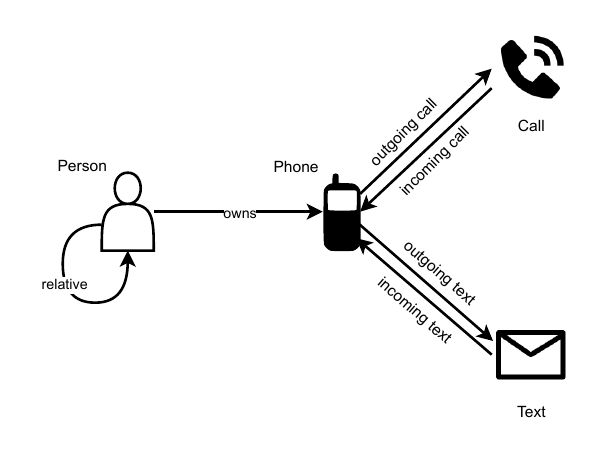}
		\caption{The ontology after reifications} 
		\label{fig:ontology-reify}
	\end{minipage}
\end{figure}

In a document-oriented store, the most common approach for modeling a knowledge graph is to represent its vertices as documents~\cite{besta2023demystifying}. A graph model is mapped to one or more document index $D$, where vertices are mapped to documents stored within those indices. More specifically, each document $d \in D$ may be mapped to a vertex $v \in V$. For instance, Alice, Bob, and Charlie are mapped to documents from a \textsf{People} index, while their phones are mapped to documents from a \textsf{Phones} index (see Fig.~\ref{fig:graph-as-docs}). Abusing the notation, we can interchangeably refer to the vertex $v$ as the document $d$ that represents it, and vice versa. 

Different approaches exist for modeling edges. Simple edges without attributes can be derived dynamically at query time: an edge $e_{vw}$ from vertex $v$ to vertex $w$ exists if values from a specific field of the document $v$ share common elements with values from another specific field of the document $w$. For instance, we know that Alice owns a certain phone number because the document representing the phone contract includes the same social security number as the document representing Alice (see Fig.~\ref{fig:graph-as-docs}). More formally, for two indices $A$ and $B$, consider fields $s$ and $t$ in their respective documents. We denote the values of field $s$ in a document $v \in A$ as $v[s]$, and similarly $w[t]$ denotes the values of field $t$ in document $w \in B$. An edge $e_{vw}$ exists if and only if there is an intersection in the values of these fields, i.e., $v[s]=w[t]$, implying at least one common value. \footnote{Under this mapping, edges can be defined through various conditions, not limited to simple equality. The graph model can accommodate more complex conditions, including range-based conditions, multi-dimensional conditions, pattern matching, etc. For clarity, we primarily use equalities in our examples.} 

Beyond this basic graph model, Siren Federate can also accommodate property graph models, where edges themselves contain properties and are treated as first-class concepts~\cite{BonifatiPGM}. In a property graph model, an edge is not merely a connection between vertices but a record with its own attributes (e.g., weight, timestamp, type). To represent such edges in a document-oriented store, we use a dedicated index $E$ for each relation type, where each document $e \in E$ corresponds to an edge with properties. Alongside its own attribute, this edge document contains fields that reference both the source and target vertices. For instance, we may index a document with Alice's and Bob's phone numbers, a date, and a time duration to represent the call between the two (see Fig.~\ref{fig:graph-as-docs}). This approach is reflected in the ontology by the introduction of an additional entity in place of the original relation, and associating the new entity with the original ones through new relations, in a process similar to reification~\cite{RDFPrimer}. For instance, the ``call relation'' could be replaced by a ``call entity'' connected to the phones via new ``outgoing call'' and ``incoming call'' relations as shown in Fig.~\ref{fig:ontology-reify}.

IR systems provide flexible data modeling and advanced search capabilities that enable searching the content of documents that encode a graph. However, they lack the relational join operations necessary for exploratory graph analysis. For instance, suppose we want to find all vertices adjacent to the vertex associated with phone number ``+353 85052***''. From the ontology, we know that we must join the documents for the ``person'', ``call'', and ``text'' entities with the document associated with that phone number -- an operation not supported by traditional IR systems. Siren Federate bridges this gap by allowing joins within the document-oriented model, thus supporting the analysis of knowledge graphs.

Several works attempt to bridge document-oriented and relational models by mapping the first into the second \cite{CLP-2013,KAA-2016,RF-2013,VFH-2017}. However, these approaches are limited to what relational databases propose and miss optimizations offered by IR systems. Siren Federate takes the opposite approach, mapping from a relational data model to a document-centric data model to fully leverage what IR systems offer: in the relational model, a join combines rows from multiple tables into a new table; in the document-oriented model, queries are applied to documents in an index and return matching documents. 

Siren Federate expresses a join operation $\bowtie$ within this model as the process of finding documents from an index (the parent set) that are related to documents from another index (the child set) according to specific conditions. Siren Federate implements
\begin{inparaenum}[(a)]
	\item the \textit{semi-join} $\ltimes$ for filtering the parent set's documents based on the child set's documents; and
	\item the \textit{inner-join} $\bowtie$ for extending the parent set's documents with fields from the matching child set's documents.
\end{inparaenum}\footnote{The Siren Federate domain-specific language for joins is documented at \url{https://docs.siren.io/siren-federate-user-guide/37/siren-federate/query-dsl.html}} 
These operators forms the foundation for implementing graph operations over our document-oriented distributed system. Specifically, an inner-join $A \bowtie_{v[s]=w[t]} B$ explicitly enumerates edges $e_{vw}$ by identifying pairs of vertices connected according to join condition equality. A semi-join $A \ltimes_{v[s]=w[t]} B$ efficiently determines which vertices $v \in A$ have outgoing edges towards vertices $w \in B$ by some edge $e_{vw}$ without enumerating all vertex pairs explicitly. Abusing the notation, in the following we will just write $A \bowtie B$ (resp. $A \ltimes B$) when the join keys are clear from the context.

\begin{figure}
	\centering
	\includegraphics[width=.6\textwidth]{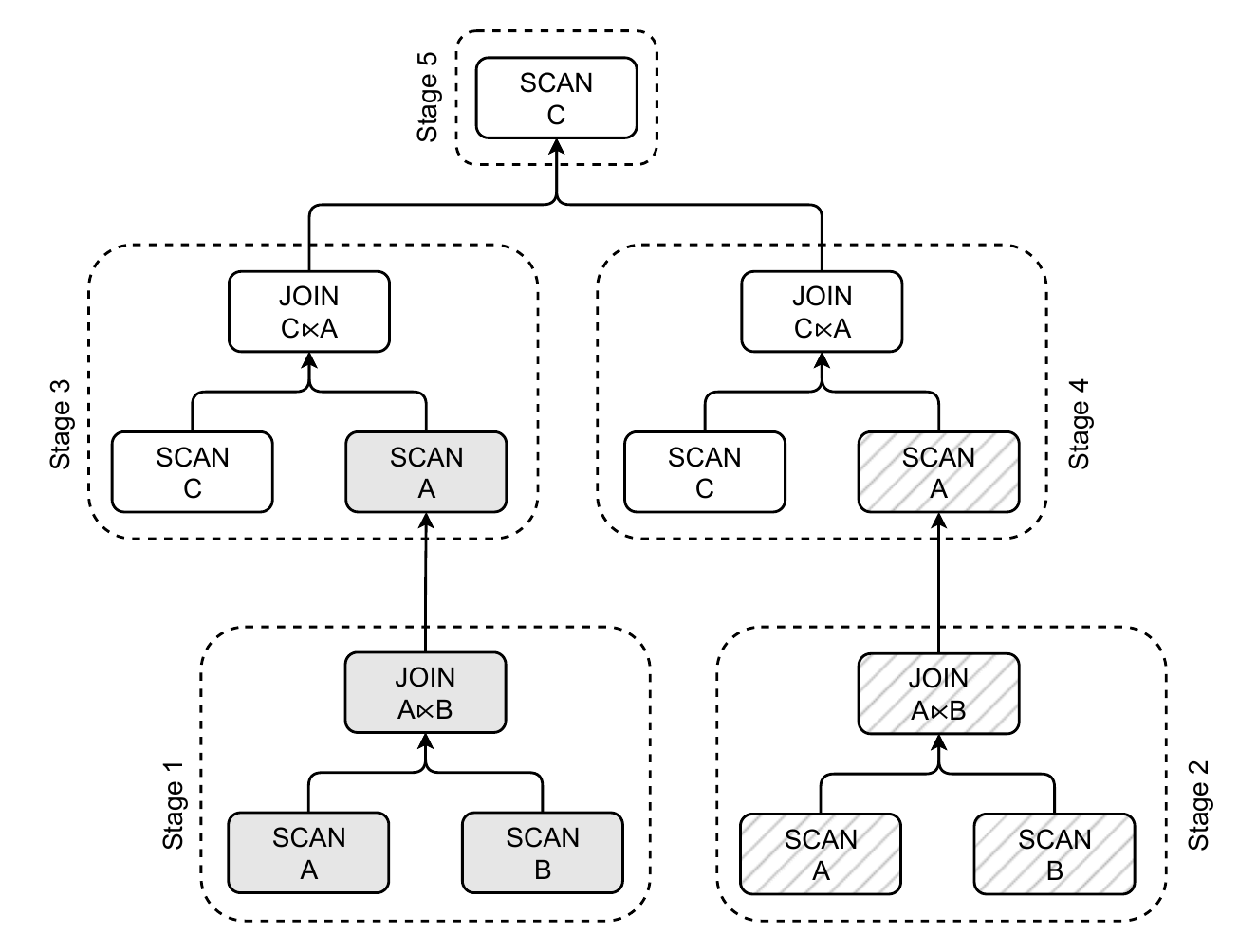}
	\caption{A staged logical query plan}
	\label{fig:query-plan}
\end{figure}

The logical steps for Siren Federate to join two document sets from indices A and B are highlighted in gray in Fig.~\ref{fig:query-plan}. Two of the steps involve a \textsf{SCAN} operation, searching over the parent set $A$ and child set $B$ to retrieve subsets of documents to be joined. These documents may need to be exchanged across the computing cluster according to one of the different strategies described in Sec.~\ref{subsec:join}. The parent and child subsets are then locally joined on the cluster's nodes by a \textsf{JOIN} operation, using data structures like hash tables, inverted indices, or k-d trees. The join results are tuples (in the relational sense) representing documents from the parent index that have fulfilled the join conditions. These tuples are then used by another \textsf{SCAN} operation to filter the parent index, retrieving the parent documents that meet the join conditions. This model also supports multi-join operations, with multiple child sets joined with the parent set. This is represented using a non-binary tree structure, where each \textsf{SCAN} operation can be associated with more than one child \textsf{JOIN} operation (see Fig.~\ref{fig:query-plan}).

Siren Federate follows a \textit{late materialization} approach, scanning only fields from parent documents to evaluate a join operation and avoid manipulating entire documents. Each document is associated with a global ID (see Sec.~\ref{subsec:log}), to uniquely identify it across the system. Tuples produced by the join operation include this ID, rather than the entire document content. Upstream operations can use this ID to materialize necessary fields. This strategy is used for various operations, such as filtering, sorting, aggregating, and retrieving document content. These operations are delegated to the underlying Elasticsearch engine, which is optimized for handling such tasks efficiently.

Given the document-centric model, tuples produced by the join must be grouped and sorted by the global document ID, as the join may produce scattered tuples about the same parent document, for example in the case of many-to-many relationships. This enables the parent \textsf{SCAN} operation to efficiently merge the join output based on ordered document IDs, which align with the natural order of the underlying log-structured storage (see Sec.~\ref{subsec:log}). We employ efficient exchange strategies for optimizing the grouping and sorting operations (see Sec.~\ref{subsec:join}).

Siren Federate uses this logical model to integrate relational joins into the document-oriented model. This representation drives the architecture design and runtime behavior of Siren Federate. For instance, the adaptive query planner uses it to stage the query plan execution. The semantic information embedded within this model is used by the semantic caching, but also for folding the query plan. Finally, this model retains the search engine's capabilities to efficiently execute filters, sorting, and aggregations.

\section{Siren Federate Architecture}\label{sec:arch}

\begin{figure}
	\centering
	\includegraphics[width=0.6\textwidth]{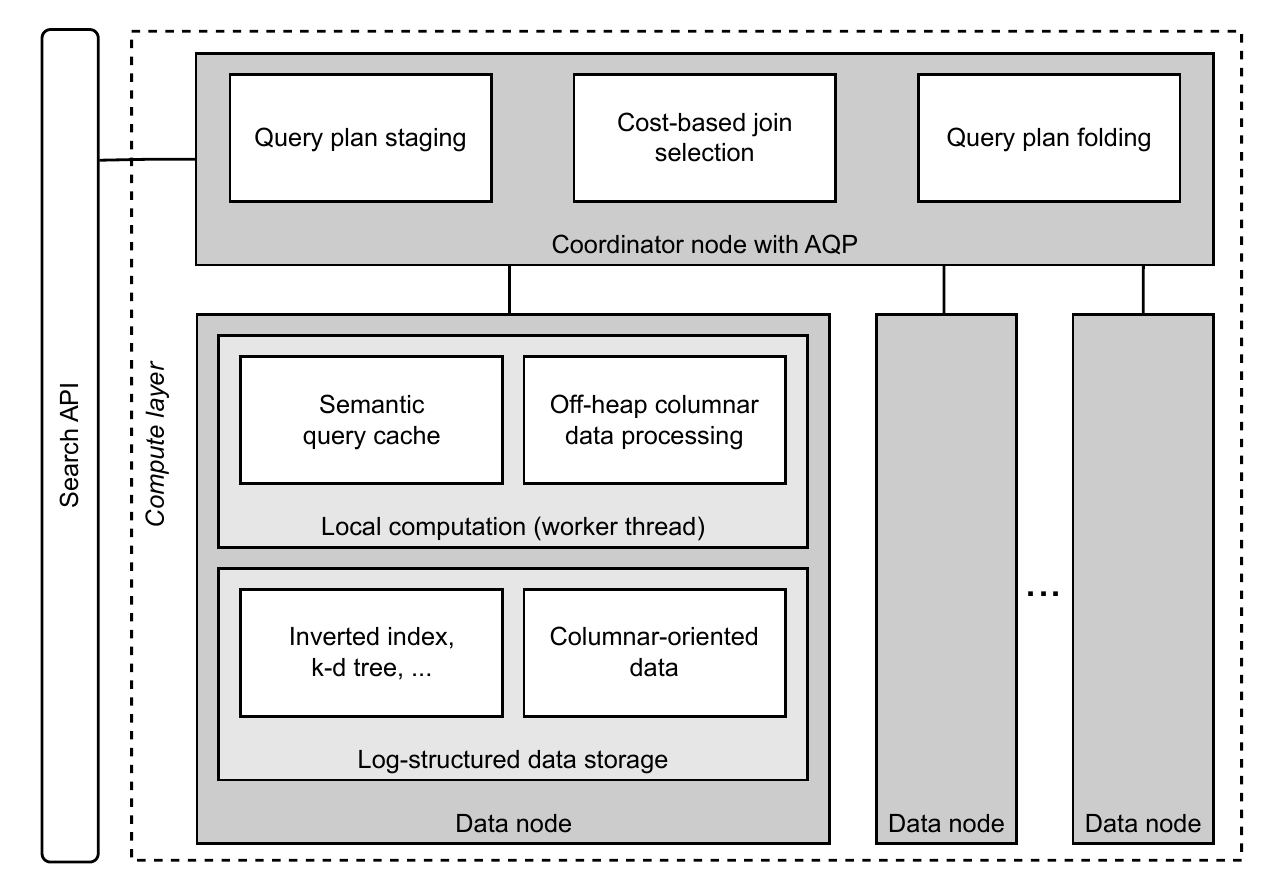}
	\caption{Siren Federate's architecture} 
	\label{fig:arch}
\end{figure}

This section introduces the core architectural components of Siren Federate, shown in Fig.~\ref{fig:arch}. Siren Federate acts as the compute layer of an investigative system, leveraging the distributed computing and storage architecture of Elasticsearch for scalability. The application layer of the investigative system relies on Siren Federate's relational and analytical capabilities via its search API.

The distributed IR system consists of a cluster of computing nodes. Each node plays a different role in the cluster: coordinator nodes are responsible for planning the execution of a request received from the search API, while data nodes are responsible for storing data and executing operations dictated by the coordinator’s query plan. This architecture ensures sub-second to seconds response time at scale, as computational load is distributed across data nodes, which can independently process the log-structured data storage to produce results (Sec.~\ref{subsec:log}). Data nodes execute scan and join operations using a columnar data processing model (Sec.~\ref{subsec:vect}) and different join algorithms (Sec.~\ref{subsec:join}). The query plan defined by the coordinator is divided into multiple stages (Sec.~\ref{subsec:aqp}). Redundant operations of the query plan are folded to avoid unnecessary computation (Sec.~\ref{subsec:fold}). The logical query plan is processed iteratively, stage-by-stage, interleaving its physical planning with its execution. At each iteration, a cost-based query optimizer checks the semantic cache to reuse existing join results (Sec.~\ref{subsec:cache}) or selects the most efficient join algorithm.

\subsection{Log-Structured Distributed Data Store}\label{subsec:log}
Siren Federate leverages Elasticsearch’s distributed data store, which horizontally partitions data across nodes using document sharding. An index is partitioned into shards, and each document is routed to a shard. A shard is a Lucene index \cite{LIA}, based on a log-structured model \cite{One+96}, and composed of one or more index segments. The log-structured model adopts an append-only update strategy and consists in creating a file-based data structure called index segment. Segments are immutable and get merged over time or when a size threshold is reached. This append-only model allows for
\begin{inparaenum}[(1)]
	\item implementing a lightweight read-lock mechanism to guarantee data consistency during the execution of distributed joins, supporting concurrent query execution with real-time data updates; and
	\item dynamically generating a global document IDs by combining shard and segment IDs with the document's insertion order, thanks to the immutability of segments.
\end{inparaenum}
This global ID enables the quick location of a document's physical position in the cluster, and late materialization as explained in Sec.~\ref{sec:bridge}.

\subsection{Columnar In-Memory Processing}\label{subsec:vect}
Siren Federate stores data for intermediate join computation into off-heap main memory using a columnar layout and leverages compression algorithms optimized for specific data types. During join operations, only necessary fields, such as join key fields and global document IDs, are processed. The data exhibits a tabular structure, with tuples corresponding to documents and columns to their fields. There are two approaches for processing tabular data: row-at-a-time and column-at-a-time.

The row-at-a-time approach reads whole tuples even if only a few columns are needed, leading to CPU cache misses and negatively impacting the performance. Following best practices from \cite{Kersten2018}, Siren Federate adopted the column-at-a-time approach, improving the query performance by a factor of 2 compared to the row-at-a-time implementation.\footnote{\url{https://info.siren.io/content/siren-benchmark-whitepaper}}

The column-at-a-time approach uses a batch-processing pipeline. Each batch stores a fixed number of tuples, stored in a columnar fashion. The size of a batch is optimized to fit within the CPU cache line to avoid cache misses. A profiling tool\footnote{\url{https://github.com/async-profiler/async-profiler}} showed an increase of the CPU cache usage with the column-at-a-time approach: Siren Federate ver.~27.5 increased ``cache-references'' by 25\% compared to ver.~22.6 that uses the row-at-a-time approach, demonstrating enhanced CPU cache utilization.

\subsection{Distributed Join Algorithms}\label{subsec:join}
\begin{figure}
	\centering
	\includegraphics[width=.95\textwidth]{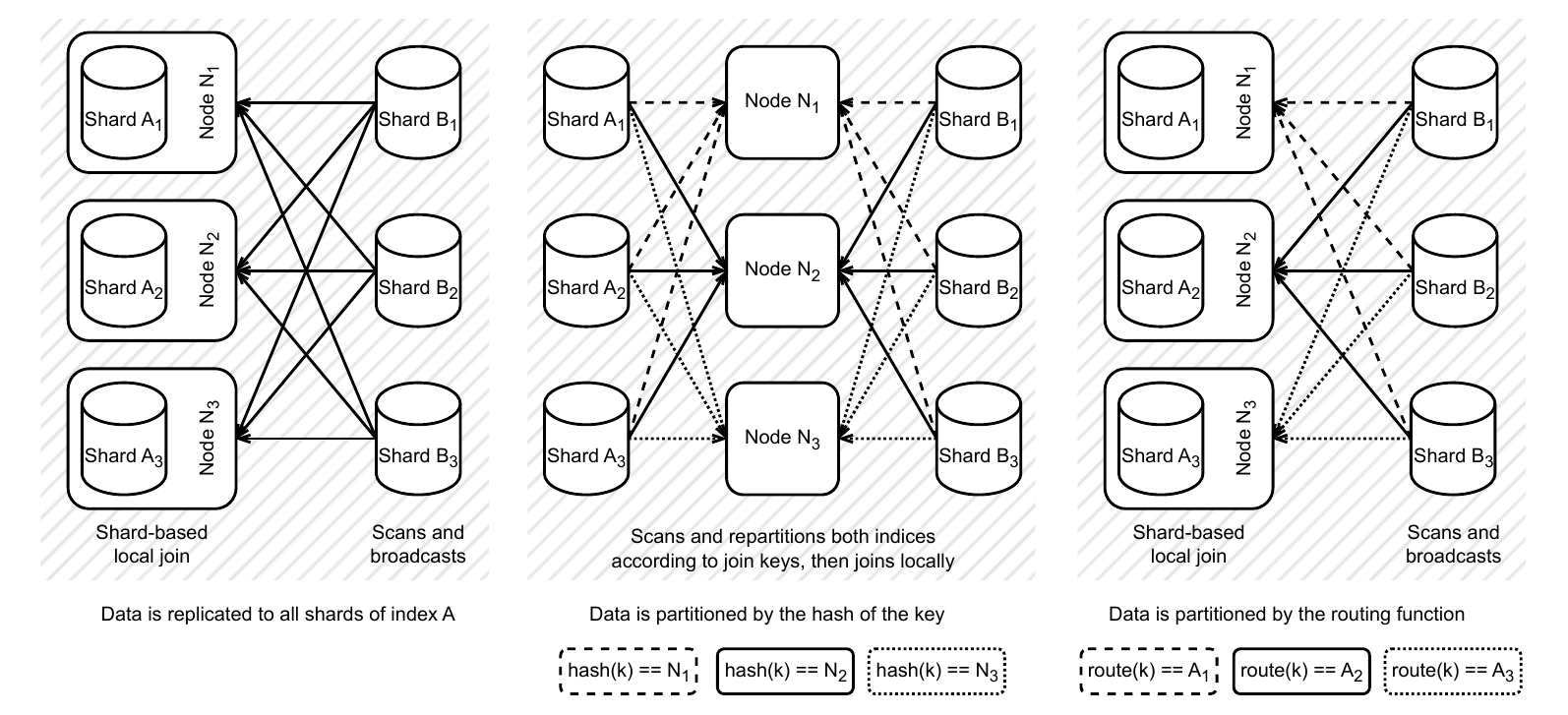}
	\caption{Federate's distributed join algorithms: (left) Broadcast Hash/Index Join, (middle) Partitioned Hash Join, (right) Routing Join. Arrows represent data exchange between computing nodes}
	\label{fig:join-algos}
\end{figure}

Siren Federate implements join techniques that leverage the intrinsic data structures of the underlying IR system to ensure scalability and high performance. An example with two distributed indices $A$ and $B$ is shown in Fig.~\ref{fig:join-algos}. Both indices are partitioned into three shards, whose data needs to be exchanged across the computing nodes in order to be joined. The available join strategies are
\begin{description}[noitemsep,style=unboxed,leftmargin=0cm]
	\item[Broadcast Hash Join] Data from the child index is forwarded to all computing nodes hosting shards of the parent index (see Fig.~\ref{fig:join-algos}, left). Local hash tables, created from the received data, are probed while scanning the parent index's columnar storage. Worker threads process segments in parallel, and local hash tables are shared across these threads.
	\item[Broadcast Index Join] This strategy utilizes Lucene's inverted indexes (akin to burst tries \cite{Heinz2002}) for binary values, and Bkd-trees \cite{Procopiuc2003} for numerical values. Data are exchanged like the broadcast hash join (see again Fig.~\ref{fig:join-algos}, left), but the child set data is used for index lookups over the parent set, eliminating exhaustive scans of the columnar storage. Worker threads process segments and probe the index with the received data. This is effective for graph expansion or path finding tasks, where the objective is to incrementally expand relationships from a group of records. Empirical assessments in Sec. \ref{sec:evaluation} show competitive performance when joining thousands of records with larger relations (in the billions).
	\item[Partitioned Hash Join] Inspired by \cite{SCD-16}, it leverages the columnar storage to scan data from the parent and child indices, partitioning data across computing nodes, and creating localized hash tables for each partition (see Fig.~\ref{fig:join-algos}, middle). This method employs morsel-driven parallelism and involves a two-step partitioning to create fixed-sized work units: an initial node partitioning at the scan level (sender side) and a second partitioning at the join level (receiver side). This method achieves better parallelism and reduced memory and network overheads compared to strategies like the broadcast hash join. In Fig.~\ref{fig:join-algos} only three computing nodes are shown for the sake of space. However, this join strategy can leverage all cluster nodes regardless the number of shards. Empirical evaluations in Sec. \ref{sec:evaluation} indicates horizontal (relative to the number of data nodes) scalability.
	\item[Routing Join] Similar to the broadcast hash join, it leverages the document sharding to reduce network traffic. It reuses the sharding routing function of the parent index to partition and exchange the child set's tuples to the corresponding parent set's shards~\cite{CCD-23} (see Fig.~\ref{fig:join-algos}, right). Each worker thread employs either a hash table-based strategy (like the broadcast hash join) or an inverted index-based strategy (like the broadcast index join) to compute the results. Preliminary experiments (not presented in this work) indicate a 30\% reduction in response times compared to the broadcast hash join strategy.
\end{description}

These strategies optimize specific scenarios. The role of the query planner (Sec~\ref{subsec:aqp}) is to select the most cost-effective join strategy by considering factors such as shard topology and set cardinality to optimize the cluster's utilization.

\subsection{Adaptive Query Planner}\label{subsec:aqp}

Accurate join cardinality estimation is crucial for planning the most effective join algorithm. This is challenging with complex query plans with deeply nested joins, common in investigative scenarios. Traditional static methods, based on histograms and cardinality estimation formulas, often yield inaccurate estimations due to assumptions like attribute independence and distribution uniformity, resulting in sub-optimal selection of join algorithms. These inaccuracies are exacerbated as the complexity of the query plan increases~\cite{Lei+17}. Index-based join sampling, while more accurate, is computationally expensive, especially in distributed systems where it requires data shuffling across the network, and also suffers from inaccuracies with long sequence of joins.

To address this, Siren Federate implements an adaptive query planner (AQP) that interleaves planning and execution via stages \cite{DZV07}. This approach collects runtime statistics during execution, allowing more accurate cardinality estimation compared to static methods, especially for long sequence of joins. This enables to dynamically adjust the query plan based on real-time feedback. AQP operates in several key phases:
\begin{description}[noitemsep,style=unboxed,leftmargin=0cm]
	\item[Logical Plan Generation] The planner generates a logical query plan divided into stages. Each corresponds to a materialization point where an intermediate result is fully created before proceeding further. Typically, it comprises a logical join and two logical scans.
	\item[Physical Optimization] The planner gathers statistical information, computes costs for various join strategies, and selects the optimal one. This is repeated for each stage, leveraging runtime cardinality estimates from already computed nested joins (stages).	
	\item[Execution] The physical sub-graph of each stage is executed, materializing intermediate results before proceeding.
	\item[Parallelization] The query plan enables parallel execution of independent stages. Independent stages are executed concurrently, while dependent stages must wait for predecessors to complete.
\end{description}

To illustrate how AQP works, consider a dataset with three indices. An index $A$ of documents representing cellphones, with fields containing the phone number, the operator, reference to the person who subscribed the contract, etc. An index $B$ of documents representing the online activity of a person such as social media posts, with fields like person's identity and textual content. An index $C$ of documents representing each call detail record (CDR) with fields such as time, duration, completion status, source and destination numbers of a call~\cite{sammons2014basics}. Imagine we want to find all CDRs related to phones used by people involved in suspicious online activities. AQP would generate a logical query plan in stages (logical plan generation) as shown in Fig.~\ref{fig:query-plan}. Assume filters (e.g., keyword matching or vector search) applied to $B$ identifies crime-related posts, then the set of ``phones used by people involved in suspicious online activities'' is the result of $A \ltimes B$ (Stages 1 and 2). The set of CDRs where these phones are the caller is returned by $C \ltimes A$ using the CDR's caller as the join key (Stage 3). Similarly, the set of CDRs where these phones are the callee is returned by $C \ltimes A$ using the CDR's callee as the join key (Stage 4). The disjunction of these two sets produces the desired results (Stage 5). Different join strategies may be used depending on the statistical information gathered from the previous stages (physical optimization and execution). Since Stages 1 and 2 are independent, they can be executed in parallel before moving to Stages 3 and 4 (parallelization).

\subsection{Query Plan Folding}\label{subsec:fold}
User queries often contain redundant operations that negatively impact query processing, such as repeated searches or joins. Redundancies commonly occur when investigating related entities through various graph topologies, boolean expressions, or batched requests targeting the same entities with diverse filters or aggregations. Redundant operations affect also SQL query processing ~\cite{Lu2023,Sahal2018}. 

To address this challenge, Siren Federate adopts a query plan folding that uses the semantic definition of query operators to detect and merge redundant operators across one or more logical query plans. The semantic definition of an operator captures its logical meaning, structure, dependencies, and state of the data tables it involves~\cite{DC-22}.

Siren Federate handles not only the folding of selection and scan operations \cite{Sahal2018}, but also of join operations. The folding strategy consolidates redundant operations into a unified shared operator. In the previous AQP example, the operators from Stage 1 and its subsequent \textsf{SCAN A} are folded with those from Stage 2 and \textsf{SCAN A} (highlighted by a hatched background in Fig.~\ref{fig:query-plan}) since they represent the same join. However, Stage 3 and 4 are not folded as they have different join conditions: the CDR's caller number is used in Stage 3, while the CDR's callee number is used in Stage 4. Similarly, \textsf{SCAN C} is not folded between the two stages because it scans different fields.

\subsection{Semantic Caching}\label{subsec:cache}
In exploratory graph analysis, iterative analysis often results in recurrent execution of the same join operations. By caching these, the system can optimize subsequent related queries, reducing the latency and computational load.

Semi-joins are well-suited for caching compared to other join types, as their outputs can be represented as sets of document IDs. Exploiting this, Siren Federate employs semantic caching, relying on the semantic definition of query operators (Sec.~\ref{subsec:fold}). Compared to conventional caching methods which operate at the query syntactic level~\cite{Pap+15}, semantic caching~\cite{DC-22} indexes cache entries according to query operator semantics, guarantying data consistency and resilience to changes in the underlying data, even when query operators depend on multiple data sources derived from descendant query operators.
Unlike traditional caching strategies that focus on raw results which can lead to large memory overhead, the semantic caching strategy uses compact bitset representations to efficiently encode semi-join outputs. This reduces memory consumption and increases the potential amount of cached operations, enhancing the overall system efficiency.

To illustrate this, take the example from Sec.~\ref{subsec:aqp}. In a subsequent iteration, the investigator wants to identify, from a new index $D$, people owning phones involved in previously found CDRs. This adds a new join $D \ltimes C$, with $C$ being the results of the previous iteration. The AQP generates a logical plan including the subtree from Fig.~\ref{fig:query-plan}. With semantic caching, the results of the subtree can be reused, meaning that only the additional join $D \ltimes C$ must be computed.

This strategy benefits graph analytics, particularly path finding algorithms which can be represented as sequences of semi-joins. Semantic caching of semi-joins reduces redundant computations, minimizing the number of operations and associated I/O, and resulting in a more efficient execution of graph queries (see Sec.~\ref{sec:sjd}).

\subsection{Principles for a competent graph analytics system}
\label{subsec:principles}

In \cite{ten2023duckpgq}, ten Wolde et al. identify eight core features necessary for a competent graph analytics system. Siren Federate leverages and extends the capabilities of Elasticsearch to meet these principles:
\begin{description}[noitemsep,style=unboxed,leftmargin=0cm,topsep=5pt]
	\item[Fast Scans on Elements with Schema] Siren Federate uses Elasticsearch's dynamic schema capabilities and column-oriented storage for fast attribute scanning. Dynamic schema-awareness offers flexibility in handling knowledge graph variability and optimizes query processing by adapting to the data structure.
	
	\item[Skippable Compressed Columnar Storage] Elasticsearch supports fast columnar scans with data skipping via pushed-down predicates, combining columnar storage ordered by document IDs with an inverted index or k-d trees.

	\item[Vectorized or Data-Centric Execution] Siren Federate adopts column-at-a-time (vectorized) processing for its data pipelines during scan and join operations, and an in-memory vector format with compression and data skipping.
	
	\item[Morsel-Driven Multi-Core] Siren Federate uses morsel-driven parallelism during scan and join operations to distribute constant-sized work units (morsels) across worker threads, reducing load imbalance and optimizing CPU cache usage. Dynamic index segment splitting is critical for better parallelism during scans of large segments, and reducing query execution latency.
	
	\item[State-of-the-Art Query Optimization] Siren Federate's AQP splits query plans into stages and employs runtime estimation to avoid the overhead of traditional table sampling, ensuring efficient query execution.
	
	\item[Bulk APIs/Algebras] Elasticsearch's boolean algebra operates as a bulk API for predicate evaluation by enabling manipulation of document sets efficiently. Siren Federate integrates relational algebra that also functions as a bulk API.
	
	\item[Out-of-Core Buffer Manager] While performing in-memory join operations, Siren Federate leverages Elasticsearch's ability to handle out-of-core data sizes efficiently during scan. Frequently accessed columns are cached at the operating system level, ensuring scan operations in RAM.
	
	\item[Explicit Control over Memory Locality] Siren Federate uses  off-heap memory management to reduce garbage collection overhead when handling gigabytes of data in memory for short durations, and optimizes memory locality through columnar storage, morsel-driven parallelism, and effective data partitioning.
\end{description}

To further enhance performance in a distributed environment, Siren Federate adopts three additional core features:
\begin{description}[noitemsep,style=unboxed,leftmargin=0cm,topsep=5pt]
	\item[Data Locality] Leveraging data locality minimizes data movement across the network, a common bottleneck as it requires additional intermediate serialization and copy of the data. Siren Federate performs late materialization of documents using global IDs to quickly locate documents in the cluster and co-locates data by reusing existing data routing coming from document sharding.

	\item[Data Exchange] Effective data exchange operators exploit the physical distribution and structure of the storage to maximize memory locality without data reorganization. Preserving the implicit order, even partially, of materialized tuples during scan and join operations improves the performance of the exchange operator, which must group and sort tuples based on the document ID as explained in Sec.~\ref{sec:bridge}. Radix partitioning \cite{Zhang2019DataPF} is a highly efficient method for clustering tuples from a range of documents together, improving document sorting.
	
	\item[Caching] Implementing compact caching strategies for intermediate results allows reuse across queries and users, reducing redundant computations and improving overall system efficiency, especially in incremental exploratory scenarios.
\end{description}

\section{Semi-Join Decomposition for Path Queries}\label{sec:sjd}

Investigative intelligence workflows frequently require querying complex chains of relationships in large-scale knowledge graphs. As discussed in Sec.~\ref{sec:investigative:requirements}, supporting these workflows requires scalable mechanisms for path-finding queries and traversals. Analysts must answer fundamental questions such as ``Are these two suspects indirectly connected?'' or ``Which entities connect them?'', often traversing long graph paths. In such scenarios, analysts must inspect the actual paths -- not just to check reachability or compute distances, but to inspect each intermediate entity or event along the way. Path enumeration therefore emerges as a central operation in investigative workflows, yet it poses significant scalability challenges in distributed systems.

The enumeration of all shortest paths between two entities -- known as the single-pair all-shortest-paths problem -- is computationally challenging due to the rapid combinatorial expansion of potential solutions. Traditional methods struggle with this complexity, as intermediate results generated during enumeration (such as expanding join tables or maintaining traversal predecessors) grow rapidly and can quickly exhaust memory resources.

This combinatorial explosion is a recurring limitation in existing path enumeration techniques, as detailed in Sec.~\ref{sec:related-work:path-query}. To address this, we introduce Semi-Join Decomposition (SJD), a technique designed for scalable, memory-efficient path enumeration.\footnote{The SJD technique was first described and patented in \cite{PTD-2022}.} SJD decomposes deep path queries into a sequence of semi-joins that iteratively prune unreachable candidates, significantly reducing intermediate result sizes. Although SJD applies broadly to general path queries, we illustrate its benefits through the all-shortest-paths use case, which shows the practical and computational challenges that SJD addresses.

In the following subsections, we formalize the reachability and path query problems (Sec.~\ref{sec:sjd:path-problem}), present the SJD method in detail (Sec.~\ref{sec:sjd:idea}), and describe its integration into Siren Federate’s distributed architecture (Sec.~\ref{sec:sjd:integration}).

\subsection{Path Queries and the Path Enumeration Challenge}\label{sec:sjd:path-problem}

At the core of graph analysis lies the path enumeration problem -- identifying and retrieving all paths that meet specified criteria. This becomes particularly challenging as the number of potential paths between nodes can grow exponentially with graph size and connectivity. This exponential growth of intermediate results is a fundamental challenge that traditional approaches struggle to address, as discussed in Sec.\ref{sec:related-work:path-query}.

Reachability and shortest-path problems highlight these enumeration challenges. In graph theory, the reachability problem consists in determining whether a node $v \in V$ can be reached from another node $u \in V$, i.e., if there exists at least one path $p$ of finite length between the two nodes. A path $p$ is an ordered list of edges $[e_{uw_1},e_{w_1w_2},\dots,e_{w_nv}]$. Equivalently, a path $p$ can be represented as an ordered list of nodes $[u, w_1, w_2, \dots, w_n, v]$ where $e_{w_iw_j} \in E$ if nodes $w_i$ and $w_j$ appear consecutively in the list (i.e., $j = i + 1$). 

The single-pair shortest path problem extends reachability by finding all paths between nodes $u,v \in V$ having the minimum possible length in terms of edge count. This corresponds to the ``\textsf{ALL SHORTEST}'' semantics in modern graph query languages such as GQL~\cite{deutsch2022graph}. Solutions to the shortest path problem also solve reachability, as any shortest path confirms that one node is reachable from the other.

For practical efficiency, path queries are often bounded with a maximum length $L$. This constrains the reachability problem to test if $v$ can be reached from $u$ following a path with at most $L$ edges, and limits the shortest path problem to paths whose length is at most $L$. However, even with these constraints, the computational complexity remains challenging in large graphs as it grows with the number of vertices and edges~\cite{cormen}.

The difficulty of path enumeration becomes apparent when we consider the growth pattern of potential paths. Let $b$ denote the average branching factor (out-degree) of nodes in a graph. For each additional hop in a path query, the number of potential paths from a node multiplies by approximately $b$. Thus, for a path of length $L$, there could be up to $O(b^L)$ distinct paths to consider -- an exponential growth that quickly overwhelms naive approaches~\cite{modernAI}. While determining the mere existence of a path or computing a single shortest path can be done efficiently, enumerating all paths meeting specific criteria becomes exponentially more difficult as path length increases.

Beyond shortest paths, many path queries involve bounded path enumeration with constraints. These might include finding all paths between nodes that are at most $L$ hops long, or paths that satisfy specific conditions on intermediate nodes or edges (e.g., ``only through legal entities in a given country''). Even with such constraints, the potential number of paths to enumerate remains exponentially large in densely connected graphs.

\subsection{Semi-Join Decomposition}\label{sec:sjd:idea}

To overcome the limitations of traditional BFS and join-based approaches, we introduce a technique called Semi-Join Decomposition (SJD). This query planning technique avoids intermediate path materialization by leveraging semi-joins, decomposing complex chains of inner-join into multiple simpler queries, significantly mitigating the exponential explosion of intermediate data during multi-hop graph traversal.

The central idea of SJD is that many intermediate paths generated by traditional methods are ultimately discarded -- not because they do not match join conditions at each hop, but because they fail to reach the target and do not lead to complete paths. Rather than fully constructing these paths only to prune them later, SJD incrementally eliminates candidates using a series of semi-joins by propagating reachability constraints.

\paragraph{Core Idea}

The essence of semi-join decomposition lies in replacing a chain of inner-joins used for a single-pair shortest path query with multiple queries, each structured as sequences of semi-joins. Consider the original inner-join chain for a path of length $l$, i.e., $D_1\bowtie D_2 \bowtie \dots \bowtie D_{l+1}$. This is decomposed into $l+1$ queries $q_k$ (for $k=1,\dots,l+1$), each consisting of a sequence of semi-joins and targeting a specific ``layer'' of vertices along the desired path. Each decomposition query $q_k$ returns a set of documents $R_{q_k}$ that correspond to vertices appearing at position $k$ in at least one path from the source $u$ to the target $v$. In other words, each document $w \in R_{q_k}$ belongs to at least a path $p = [u=w_1, \dots, v=w_{l+1}]$ and appears precisely at position $k$.

For example, given a path of length $3$ across indices $\{D_1, D_2, D_3, D_4\}$, the inner-join sequence $D_1\bowtie D_2 \bowtie D_3 \bowtie D_4$ can be decomposed into four semi-join queries:
\begin{inparaenum}[(1)]
	\item $q_1 = D_1 \ltimes D_2 \ltimes D_3 \ltimes D_4$;
	\item $q_2 = (D_2 \ltimes D_1) \land (D_2 \ltimes D_3 \ltimes D_4)$;
	\item $q_3 = (D_3 \ltimes D_2 \ltimes D_1) \land (D_3 \ltimes D_4)$;
	\item $q_4 = D_4 \ltimes D_3 \ltimes D_2 \ltimes D_1$.
\end{inparaenum}

For each tested path length $l$, the semi-join decompositions $q_k$ can be executed sequentially or in parallel. However, one query is better used as an initial reachability test, and any decomposition query $q_k$ can be used for this. For instance, the midpoint $q_{\lceil \frac{l+1}{2} \rceil}$ could be tested using a bisecting approach, and $q_{l+1}$ can be used to exploit the cached results of the reachability test for $l-1$. If this initial query returns an empty result set, there is no path of the given length, thus preventing unnecessary computation of subsequent queries. Otherwise, the remaining queries are executed and their result sets stored.

This decomposition aggressively prunes irrelevant paths early because semi-joins do not enumerate all possible combinations explicitly. Instead, they filter vertices by simply checking for the existence of matches, limiting intermediate results to a size linear in the number of vertices rather than exponential with respect to branching factors.

\paragraph{Path Materialization}

Once the relevant documents at each layer have been identified via semi-joins, full path enumeration proceeds via a guided depth-first search (DFS), constrained to documents in $\{R_{q_1}, \dots, R_{q_{l+1}}\}$:
\begin{inparaenum}[(1)]
	\item Starting from the source node, adjacent nodes are retrieved from the index using the appropriate semi-join result sets as lookup tables.
	\item Iteratively, subsequent vertices at each path position are identified via in-memory hash tables keyed by relevant field values. Each ${R_{q_k}}$ can be retrieved from the index and stored in a separate hash table with source fields as keys, enabling rapid lookups.
\end{inparaenum}

This materialization approach supports streaming delivery, allowing paths to be generated and consumed incrementally without requiring full materialization in memory. The approach also offers flexibility in how it can materialize results, supporting complete path tuples (capturing the entire path from source to destination), individual edges, or triples (similar to RDF's subject-predicate-object model).

\textit{Remark (Connection to BFS)} Classic BFS‐based approaches also organize nodes into ``layers'' but typically maintain a layered DAG of parent pointers. In our method, the sets ${R_{q_1},\dots,R_{q_{l+1}}}$ serve an analogous function, capturing exactly those vertices that appear at position $k$ in some valid path. 

\subsection{Complexity and Caching Benefits}

The size complexity of each semi-join result set $R_{q_k}$ is bounded linearly by the number of vertices $|V|$, rather than growing exponentially with the branching factor. Consequently, the combined complexity of all decomposed results remains linear with respect to $|V|$, a significant improvement over traditional sequence of joins.

However, SJD requires $l+1$ queries each involving $l$ semi-joins, i.e., $O(l^2)$ total operations versus $O(l)$ inner-joins in a naive chain. In practice, SJD remains favorable because each semi-join is significantly cheaper than a full inner join:
\begin{inparaenum}[(1)]
	\item it produces a smaller result set and exchanges less data over the network;
	\item it operates faster due to smaller memory footprints and optimized local join algorithms;
	\item it is efficiently cached using compact bitsets, as detailed in Sec.~\ref{subsec:cache}, and can be reused by subsequent decompositions.
\end{inparaenum}

Caching dramatically reduces repeated semi-join computations, as consecutive decompositions often share numerous intermediate results, allowing significant reuse of previously computed results. 
\begin{atheorem}
For each subsequent decomposition $q_i$, all semi-join operations except one are identical to those performed by the previous decomposition $q_{i-1}$, benefiting directly from cached results.
\begin{proof}
	By induction: 
\begin{inparaenum}
	\item[(Base case)] Let $q_1 = (D_1 \ltimes \dots \ltimes D_{l+1})$ and  $q_2 = (D_2 \ltimes D_1) \land (D_2 \ltimes \dots \ltimes D_{l+1})$. $(D_2 \ltimes \dots \ltimes D_{l+1})$ from $q_2$ is a sub-sequence of $(D_1 \ltimes \dots \ltimes D_{l+1})$ from $q_1$, hence it can be processed by reusing the cached results. Only $(D_2 \ltimes D_1)$ must be computed.
	\item[(Step case)] Let $q_i = (D_i \ltimes \dots \ltimes D_1) \land (D_i \ltimes \dots \ltimes D_{l+1})$ and $q_{i+1} = (D_{i+1} \ltimes \dots \ltimes D_1) \land (D_{i+1} \ltimes \dots \ltimes D_{l+1})$. $(D_{i+1} \ltimes \dots \ltimes D_{l+1})$ from $q_{i+1}$ is a sub-sequence of $(D_i \ltimes \dots \ltimes D_{l+1})$ from $q_i$, therefore it can be processed by reusing the cached results. $(D_{i+1} \ltimes \dots \ltimes D_1)$ from $q_{i+1}$ contains sub-sequence $(D_i \ltimes \dots \ltimes D_1)$ from $q_i$, therefore only the outermost join from $(D_{i+1} \ltimes (D_i \ltimes \dots \ltimes D_{l+1}))$ must be computed while the rest can be processed by reusing the cached results.
\end{inparaenum}
\end{proof}
\end{atheorem}
For instance, in the example decomposition $q_1 = D_1 \ltimes D_2 \ltimes D_3 \ltimes D_4$, computation $D_2 \ltimes D_3 \ltimes D_4$ is reused by subsequent query $q_2$ without additional processing. Even though we conceptually have $O(l^2)$ semi-joins, most are skipped by cache hits, significantly reducing the computational overhead. After caching, at most $2l$ semi-join operations need execution for a single path of fixed length $l$, thus achieving linear complexity.

When testing all path lengths from $1$ up to $L$ to find the minimal length at which the target is reachable, the difference between naive chains of joins and SJD is even more pronounced. The chains of joins will execute $l$ inner-joins for each length $l = 1, 2, \dots, L$ which amounts to a total of $O(L^2)$. In comparison, SJD will execute one semi-join query with $l$ semi-joins for each length $l$, reusing cached results almost entirely, until the reachability test is positive. Empirically, the reachability test for each new length often requires computing at most two semi-joins, while reusing the cached result sets of the other semi joins that were computed at the previous iterations. Therefore, the number of semi joins to perform the reachability tests until a path length of $l$ is found is $2\cdot(l-1)$ semi-joins. Once the reachability test is positive, the $(l-1)$ remaining semi-join queries are executed by computing only one semi-join operation (reusing the cached results for the others) for each query. This leads to an overall cost of about $3\cdot(L-1)$ semi-joins, i.e., linear in $L$. Although the exact break-even point varies, once $L$ grows past 5 or 6, SJD plus caching can require fewer total joins than the naive chains of inner-joins method.

Overall, SJD’s complexity in terms of semi-join operations scales linearly with the path length $l$ once caching is factored in, while intermediate data remain linear in $|V|$. By separating ``reachability pruning'' (through semi-joins) from ``path materialization'' (through a final DFS), SJD provides a scalable strategy for iterative path queries that avoids prohibitive computational or memory overhead.

\subsection{Integration with Siren Federate}
\label{sec:sjd:integration}

The semi-join decomposition approach integrates seamlessly with Siren Federate's architecture and optimization strategies. This integration enhances Siren Federate's ability to handle complex graph queries efficiently by leveraging several key features:
\begin{description}[noitemsep,style=unboxed,leftmargin=0cm]
	\item[Adaptive Query Planning (AQP)] Siren Federate employs an adaptive query planner (Sec.~\ref{subsec:aqp}) that iteratively optimizes query execution plans based on runtime statistics. Semi-join decomposition queries naturally fit this model, as each query step provides valuable intermediate statistical information used to choose the most efficient semi-join strategy (Sec.~\ref{subsec:join}).
	
	\item[Late Materialization and Columnar Storage] Consistent with Siren Federate's late materialization strategy (Sec.~\ref{subsec:log}), semi-join decomposition primarily manipulates compact document identifiers rather than entire document contents. Combined with Siren Federate's columnar storage and vectorized data processing (Sec.~\ref{subsec:vect}), semi-join operations access only minimal field subsets required for join computations, significantly lowering I/O and memory overhead.

	\item[Semantic Caching] The effectiveness of semi-join decomposition is substantially enhanced by Siren Federate's semantic caching mechanism (Sec.~\ref{subsec:cache}). Semi-join results are compactly represented as bitsets, which can be cached efficiently. Since multi-hop investigative queries typically involve iterative and incremental expansions, cached semi-join results dramatically reduce redundant computations across query iterations.

	\item[Join Exchange Operators and Data Locality] Semi-join decomposition exploits Siren Federate's optimized data exchange operators (Sec.~\ref{subsec:join}). These operators efficiently handle data transfers between cluster nodes, taking advantage of existing data locality and document-sharding mechanisms to minimize network traffic in highly distributed environments.
\end{description}

In Sec.~\ref{sec:evaluation}, we evaluate semi-join decomposition on complex path queries, showing that it outperforms naive join approaches as data scale increases, while maintaining interactive response times suitable for investigative workflows.

\section{Evaluation}\label{sec:evaluation}

In this section, we present an experimental evaluation of Siren Federate that focuses on two complementary aspects of the system:
\begin{inparaenum}[(1)]
	\item scalability with large data volumes in a distributed environment; and
	\item performance on complex graph querying patterns.
\end{inparaenum}
Our evaluation methodology reflects the system's primary design goals -- enabling interactive exploration of large-scale heterogeneous knowledge graphs with sub-second to second response times.

Comprehensive cross-system comparisons present practical challenges: configuring and optimizing multiple systems for fair comparison requires significant resources, particularly when considering commercial systems with potential benchmarking restrictions. We have therefore focused on providing an objective assessment of our system's performance characteristics and the impact of key architectural decisions. This approach aligns with our primary contribution of sharing practical architectural insights gained through long-term system development in an industrial setting.

Rather than comparing against multiple systems with different architectural foundations, we focus on Siren Federate's performance across critical dimensions aligned with investigative intelligence requirements: data scale (billions of documents), query complexity (from simple relation traversals to complex path finding), and concurrent usage patterns. This approach validates the effectiveness of the architectural choices and optimization techniques described in previous sections. Future work could explore more targeted comparative evaluations of specific components such as the adaptive query planner and semi-join decomposition under varying workloads and data distributions.

In Sec.~\ref{sec:evaluation:large-scale-benchmark}, we evaluate the system's performance and scalability using a synthetic dataset representing cell phone location records -- a common data source in investigative scenarios. In Sec.~\ref{sec:evaluation:ldbc}, we assess the system's graph querying capabilities using the LDBC Financial Benchmark (Finbench), analyzing how our adaptive query planning and semi-join decomposition techniques perform on standardized graph patterns. Finally, Sec.~\ref{sec:evaluation:real-world} draws insights from real-world deployments, demonstrating how these techniques translate into operational impact in production environments.

\subsection{Large Scale Benchmark}\label{sec:evaluation:large-scale-benchmark}

This section demonstrates Siren Federate's scalability and performance under realistic investigative intelligence workloads. We evaluate the system's ability to process semi-joins efficiently across three dimensions that reflect real-world deployment scenarios: computing cluster size, concurrent user load, and data volume. We focus on semi-joins as they are more suitable than inner-joins for large datasets and form the foundation of exploratory graph analysis tasks like set-to-set navigation, graph expansion, and pathfinding. Our experimental design uses a synthetic dataset modeling cellular location tracking, representing a common scenario where analysts must correlate location data across time periods and geographic boundaries. 

\paragraph{Dataset.}

We use a synthetic dataset\footnote{\url{https://gist.github.com/scampi/07e7bd556fe016a5cba6c092c3f418fb}} with 15.6 billion documents tracking the positional information of cell phones to mimic scenarios where analysts monitor phone calls. This dataset covers 100 days, with 156 million documents per day, 6.5 million unique phone identifiers, and 2,400 positions per phone. One Elasticsearch index per day is created, with 8 primary shards with no replicas. The total size of the data is 2.7~TB.

\paragraph{Setup.}

The benchmark tests use a varying number of machines with 16 CPUs, a local NVME drive for Elasticsearch data, a gp2 drive for the OS, and a 10 Gbps network link. Elasticsearch is configured with 30 GB heap memory, and Siren Federate with 16 GB off-heap memory.

\paragraph{Experiments.}

We evaluate Siren Federate with varying cluster sizes (12 to 36 nodes), data volumes, and concurrent users (1, 5 and 10). The system is setup to serve the maximum number of concurrent users. We use the following queries with different complexity:

\begin{inparaenum}[\bfseries{Q}1]
	\item joins phone numbers in a given area on one day with those in another area on another day (78 million documents per set);
	\item is similar to Q1 but over a week (546 million documents per set);
	\item given a phone number, finds other phones at the same location over 90 days (14 billion documents filtered with 2,160 documents);
	\item finds phones at the same location on two different days (156 million documents each);
	\item is similar to Q4 but over a week (more than 1 billion documents per set);
	\item is similar to Q4 but over two weeks (more than 2 billion documents per set).
\end{inparaenum}

Queries use the \emph{partitioned hash join}, except Q3, which uses the \emph{broadcast index join} due to the small cardinality of its child set. We measure the execution time for a randomly-selected query with a fixed number of concurrent users, bypassing query caches. The benchmark runs until at least 100 measurements per query are produced, reporting the 90th percentile processing time (P90).

\paragraph{Results.}

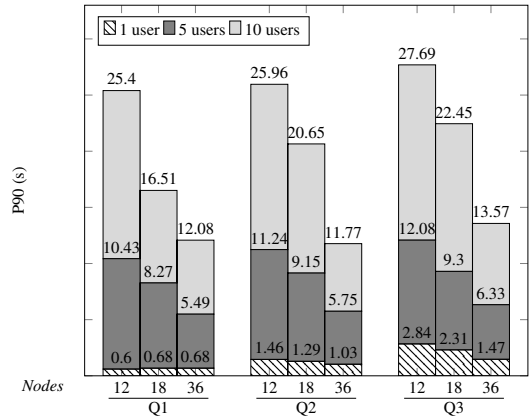
\begin{figure}
	\centering
\pgfplotstableread{%
X   Nodes  Query  Time_1user  Time_5users  Time_5users_adjusted Time_10users_adjusted Time_10users
0   12     Q1     0.60        10.43        9.83                 14.97                 25.40
1   18     Q1     0.68        8.27         7.59                 8.24                  16.51
2   36     Q1     0.68        5.49         4.81                 6.59                  12.08
4   12     Q2     1.46        11.24        9.78                 14.72                 25.96
5   18     Q2     1.29        9.15         7.86                 11.50                 20.65
6   36     Q2     1.03        5.75         4.72                 6.02                  11.77
8   12     Q3     2.84        12.08        9.24                 15.61                 27.69
9   18     Q3     2.31        9.30         6.99                 13.15                 22.45
10  36     Q3     1.47        6.33         4.86                 7.24                  13.57
}\datatable%
\newcounter{groupcount}
\pgfplotsset{width=10cm}
\pgfplotsset{%
    draw group line/.style n args={5}{%
        after end axis/.append code={%
            \setcounter{groupcount}{0}
            \pgfplotstableforeachcolumnelement{#1}\of\datatable\as\cell{%
                \def\temp{#2}
                \ifx\temp\cell
                    \ifnum\thegroupcount=0
                        \stepcounter{groupcount}
                        \pgfplotstablegetelem{\pgfplotstablerow}{X}\of\datatable
                        \coordinate [yshift=#4] (startgroup) at (axis cs:\pgfplotsretval, 0);
                    \else
                        \pgfplotstablegetelem{\pgfplotstablerow}{X}\of\datatable
                        \coordinate [yshift=#4] (endgroup) at (axis cs:\pgfplotsretval, 0);
                    \fi
                \fi
            }
            \ifnum\thegroupcount=1
                        \setcounter{groupcount}{0}
                        \draw [shorten >=-#5, shorten <=-#5] (startgroup) -- node [anchor=base, yshift=-2ex] {#3} (endgroup);
            \fi
        }
    }
}%
\begin{tikzpicture}[scale=0.7]
\begin{axis}[
    ymin=0, ymax=33,
    ybar stacked,
    bar width=20,
    nodes near coords,
    legend style={
            legend pos=north west,
            legend columns=3
    },
    xtick=data,
    xticklabels from table={\datatable}{Nodes},
    yticklabel=\empty,
    ylabel=P90 (s),
    draw group line={Query}{Q1}{Q1}{-3ex}{7pt},
    draw group line={Query}{Q2}{Q2}{-3ex}{7pt},
    draw group line={Query}{Q3}{Q3}{-3ex}{7pt},
    after end axis/.append code={
            \path [anchor=base east, yshift=0.5ex] (rel axis cs:0,0) node [yshift=-2.5ex,xshift=-2ex] {\emph{Nodes}};
    }
]
\addplot [pattern=north west lines] table [x=X, y=Time_1user] {\datatable};
\addplot [fill=gray] table [x=X, y=Time_5users_adjusted] {\datatable};
\addplot [fill=gray!30!white] table [x=X, y=Time_10users_adjusted] {\datatable};
\legend{1 user,5 users,10 users}
\end{axis}
\end{tikzpicture}
	\caption{Query times for queries Q1 to Q3 with varying number of nodes and users}
	\label{fig:results_standard}
\end{figure}

Fig.~\ref{fig:results_standard} reports P90 for queries Q1 to Q3. With one concurrent user, Q1 joins 156 million documents in subsecond time. P90 does not decrease as the number of nodes increases because the amount of data joined is too small. The latency of the join phase represents an insignificant part of the response time, while the scan phase is tied to a limited number of shards and cannot be further distributed. 

Q2 joins over 1 billion documents ($\times7$ more data than Q1), with P90 increasing by at most $\times2.5$, indicating a better usage of the computing resources. However, response time does not decrease as expected with more nodes. In fact, increasing the number of nodes by $\times3$ only decreases P90 by 30\%. This suggests that Q1 and Q2 latency is dominated by fixed overhead during the query planning and pipeline execution. This requires further analysis.

Q3, which filters 14 billion documents using the \emph{broadcast index join} strategy, shows P90 decreasing by $\times2$ as nodes increases by $\times3$. This strategy avoids partitioning and shuffling such a large parent set while still distributing load across the full cluster as the number of nodes increases.

With 5 concurrent users, P90 of Q1 increases by $\times12.5$ -- on average across cluster's configurations -- when compared to response times with 1 user; Q2 increases by $\times6.8$; Q3 by $\times4.2$. With 10 concurrent users, the increase of P90 is by at most $\times2.3$ w.r.t. 5 users. Differently than moving from 1 to 5 concurrent users, doubling the number of users doubles the response times in this case. This significant increase in latency is particularly evident for Q1, and it may underline as previously noted an overhead in query planning and execution for short queries. Such overhead may impact scaling, and needs to be investigated in future work. Nonetheless, these results show that Siren Federate scales well with the number of users and nodes, achieving subsecond to second response times over large datasets.

For Q4, Q5, and Q6, the aim is to further assess the ability of the system to scale with the amount of data, joining parent and child sets containing hundreds of millions to billions of documents each. With 36 nodes, the reported P90 is $1.92$, $5.00$, and $7.58$ seconds for Q4, Q5, and Q6 respectively. We can observe a sub-linear scaling factor with the size of the join operation. Between Q4 and Q5 the size of the join operation increases by $\times7$, but P90 only increases by 2.2 seconds. Between Q5 and Q6, the size of the join operation increases by $\times2$ but P90 only increases by 1.5 seconds. These results confirm that Siren Federate scales well with the amount of data.

\subsection{LDBC Finbench}\label{sec:evaluation:ldbc}

This section evaluates Siren Federate's graph querying capabilities using standardized, complex query patterns derived from the LDBC Financial Benchmark (Finbench). These experiments assess our query planning strategies and semi-join decomposition (SJD) technique on realistic graph traversal workloads.Using an industry-standard benchmark provides objective measurements of our system's ability to handle diverse graph query patterns -- from multi-hop traversals to shortest path computations -- that are representative of investigative workflows in financial compliance and fraud detection scenarios.

Our evaluation focuses on two key aspects of graph query performance:
\begin{inparaenum}[(1)]
\item comparing the efficiency of adaptive versus static query planning approaches across varying complexity levels and resource configurations; and
\item demonstrating the practical benefits of semi-join decomposition for path queries compared to traditional sequences of inner joins.
\end{inparaenum}
For path queries, while theoretical comparisons to BFS-based approaches are provided in Sec.\ref{sec:related-work:path-query}, our experimental evaluation compares SJD against the sequence of inner joins approach implemented within our system. This comparison allows us to measure the performance improvements of our contribution while maintaining implementation feasibility within our research scope.

These measurements validate that our architectural decisions enable performant graph analytics on document-oriented data stores, bridging the gap between rich IR capabilities and complex graph traversals. The results provide clear evidence of SJD's practical benefits, while establishing a foundation for future comparative evaluations against additional graph processing techniques.

\paragraph{Dataset.}

We use the LDBC Financial Benchmark dataset~\cite{qi2023ldbc}, denoted as \emph{Finbench}. This benchmark represents financial industry data and query patterns. Its schema consists of 5 entity and 13 edge types, e.g., persons and accounts along with money transfers. This benchmark\footnote{\url{https://ldbcouncil.org/benchmarks/finbench/}} provides pre-generated datasets of different sizes; we used the dataset with a scale factor of 10, containing 138,305,500 triples which is in terms of documents indexed 5,376,981 entity instances and 26,633,151 relations.

\paragraph{Setup.}

Since the system's performance in a distributed environment has already been assessed in Sec.~\ref{sec:evaluation:large-scale-benchmark}, this section focuses on evaluating the efficiency of local processing and optimization techniques for graph queries. Each entity and edge type from the dataset is indexed into its own index -- as described with the reification approach in Section~\ref{sec:bridge} -- in a single-node Elasticsearch cluster with index replication disabled. The computing instance is equipped with 64GB of main memory, while we conduct tests with varying numbers of CPUs: 8, 16, 32, and 64.

\paragraph{Experiments.}

We use a subset of the \textit{complex-read queries} from the Finbench \textit{transaction workload}; these queries require matching graph patterns of varying complexity. To process them, we express the queries in GQL~\cite{deutsch2022graph}. Siren Federate can interpret GQL queries and internally translate them into Federate query plans for processing. Noticeably, queries TCR1, TCR2, TCR7, TCR9, TCR11, and TCR12 require retrieving some entities’ attributes, whereas queries TCR3 and TCR5 involve computing certain paths (see Appendix~\ref{app:ldbc-gql-queries}). Due to current system limitations, we were unable to process some other queries from the dataset. Specifically:
\begin{inparaenum}[(a)]
\item TCR4 requires support for cycles;
\item TCR6 requires support for conditions on the number of connected documents;
\item TCR8 requires the graph engine to handle complex patterns, such as edges defined as the disjunction of different labels or referencing values from other nodes in a \textsf{WHERE} clause; and
\item TCR10 requires computing the Jaccard similarity between two sets.
\end{inparaenum}
These features are not currently available, but we plan to introduce them in future work.

We evaluate Siren Federate by simulating a single user or 10 concurrent users issuing these queries. Additionally, we experiment with query planning using a static or adaptive planner to assess which approach is the most efficient. For each join in the query, the static query planner (SQP) selects the most appropriate join strategy prior to the query processing and leveraging simple simple statistics estimation. On the contrary, the adaptive query planner (AQP) interleaves query planning and processing to collect and leverage more accurate runtime statistics, as described in Sec.~\ref{subsec:aqp}.

\paragraph{Results.}

\begin{figure}
	\centering

\pgfplotstableread{%
X CPUs  AQP_1user  SQP_1user  AQP_10users  SQP_10users
1 8     1428.80    1779.50    25453.70     26001.40
2 16    666.70     945.80     9287.70      9994.30
3 32    685.00     611.00     1655.90      2311.80
4 64    867.60     835.90     1256.00      1235.90
}\TCRone%
\pgfplotstableread{%
X CPUs  AQP_1user  SQP_1user  AQP_10users  SQP_10users
1 8     818.30     1199.00    22193.70     24126.60
2 16    732.00     798.90     8676.20      10071.60
3 32    727.90     594.70     1760.90      1635.00
4 64    909.80     644.80     1397.00      1131.90
}\TCRtwo%
\pgfplotstableread{%
X CPUs  AQP_1user  SQP_1user  AQP_10users  SQP_10users
1 8     6775.20    7551.80    25838.80     28029.80
2 16    6521.00    7344.90    17073.90     16677.90
3 32    5732.50    5271.00    10057.90     9837.90
4 64    6756.60    6184.20    9061.80      8712.80
}\TCRseven%
\pgfplotstableread{%
X CPUs  AQP_1user  SQP_1user  AQP_10users  SQP_10users
1 8     5263.50    5690.60    14864.20     15520.80
2 16    4983.00    5244.00    11452.40     11782.80
3 32    4171.30    3627.90    7046.90      7270.90
4 64    4742.80    4181.50    6374.90      5941.00
}\TCRnine%
\pgfplotstableread{%
X CPUs  AQP_1user  SQP_1user  AQP_10users  SQP_10users
1 8     200.10     337.00     12905.50     14210.30
2 16    140.00     325.90     5050.80      5837.90
3 32    132.90     288.90     911.70       1661.50
4 64    133.00     287.50     280.90       499.00
}\TCReleven%
\pgfplotstableread{%
X CPUs  AQP_1user  SQP_1user  AQP_10users  SQP_10users
1 8     244.90     347.90     8276.40      8943.40
2 16    223.00     241.90     3074.90      3554.80
3 32    221.00     179.00     667.00       1289.60
4 64    298.00     209.90     426.00       369.00
}\TCRtwelve%
\begin{subcaptionblock}{.45\textwidth}
	\centering
	\begin{tikzpicture}[scale=0.7]
		\begin{axis}[
			ybar,
			ymode=log,
			ymin=100,
			ymax=100000,
			enlarge x limits=0.2,
			bar width=7,
			xtick=data,
			xticklabels from table={\TCRone}{CPUs},
			grid,
			legend style={
				legend columns=2,
				font=\footnotesize
			},
			legend cell align={left},
			legend image code/.code={
				\draw [#1] (-0.125cm,-0.125cm) rectangle (0.125cm,0.125cm);
			}
			]
			
			\addplot [pattern=north west lines] table [y=AQP_1user, x=X] {\TCRone};
			\addplot [fill=lightgray] table [y=SQP_1user, x=X] {\TCRone};
			
			\addplot [pattern=north east lines] table [y=AQP_10users, x=X] {\TCRone};
			\addplot [fill=gray] table [y=SQP_10users, x=X] {\TCRone};
			
			\legend{AQP (1 user), SQP (1 user), AQP (10 users), SQP (10 users)}
		\end{axis}
	\end{tikzpicture}
	\caption{TCR1}
	\label{fig:ldbc:tcr1}
\end{subcaptionblock}
\begin{subcaptionblock}{.45\textwidth}
	\centering
	\begin{tikzpicture}[scale=0.7]
		\begin{axis}[
			ybar,
			ymode=log,
			ymin=100,
			ymax=100000,
			enlarge x limits=0.2,
			bar width=7,
			xtick=data,
			xticklabels from table={\TCRtwo}{CPUs},
			grid,
			legend style={
				legend columns=2,
				font=\footnotesize
			},
			legend cell align={left},
			legend image code/.code={
				\draw [#1] (-0.125cm,-0.125cm) rectangle (0.125cm,0.125cm);
			}
			]
			
			\addplot [pattern=north west lines] table [y=AQP_1user, x=X] {\TCRtwo};
			\addplot [fill=lightgray] table [y=SQP_1user, x=X] {\TCRtwo};
			
			\addplot [pattern=north east lines] table [y=AQP_10users, x=X] {\TCRtwo};
			\addplot [fill=gray] table [y=SQP_10users, x=X] {\TCRtwo};
			
			\legend{AQP (1 user), SQP (1 user), AQP (10 users), SQP (10 users)}
		\end{axis}
	\end{tikzpicture}
	\caption{TCR2}
	\label{fig:ldbc:tcr2}
\end{subcaptionblock}
\\
\begin{subcaptionblock}{.45\textwidth}
	\centering
	\begin{tikzpicture}[scale=0.7]
		\begin{axis}[
			ybar,
			ymode=log,
			ymin=100,
			ymax=100000,
			enlarge x limits=0.2,
			bar width=7,
			xtick=data,
			xticklabels from table={\TCRseven}{CPUs},
			grid,
			legend style={
				legend columns=2,
				font=\footnotesize
			},
			legend cell align={left},
			legend image code/.code={
				\draw [#1] (-0.125cm,-0.125cm) rectangle (0.125cm,0.125cm);
			}
			]
			
			\addplot [pattern=north west lines] table [y=AQP_1user, x=X] {\TCRseven};
			\addplot [fill=lightgray] table [y=SQP_1user, x=X] {\TCRseven};
			
			\addplot [pattern=north east lines] table [y=AQP_10users, x=X] {\TCRseven};
			\addplot [fill=gray] table [y=SQP_10users, x=X] {\TCRseven};
			
			\legend{AQP (1 user), SQP (1 user), AQP (10 users), SQP (10 users)}
		\end{axis}
	\end{tikzpicture}
	\caption{TCR7}
	\label{fig:ldbc:tcr7}
\end{subcaptionblock}
\begin{subcaptionblock}{.45\textwidth}
	\centering
	\begin{tikzpicture}[scale=0.7]
		\begin{axis}[
			ybar,
			ymode=log,
			ymin=100,
			ymax=100000,
			enlarge x limits=0.2,
			bar width=7,
			xtick=data,
			xticklabels from table={\TCRnine}{CPUs},
			grid,
			legend style={
				legend columns=2,
				font=\footnotesize
			},
			legend cell align={left},
			legend image code/.code={
				\draw [#1] (-0.125cm,-0.125cm) rectangle (0.125cm,0.125cm);
			}
			]
			
			\addplot [pattern=north west lines] table [y=AQP_1user, x=X] {\TCRnine};
			\addplot [fill=lightgray] table [y=SQP_1user, x=X] {\TCRnine};
			
			\addplot [pattern=north east lines] table [y=AQP_10users, x=X] {\TCRnine};
			\addplot [fill=gray] table [y=SQP_10users, x=X] {\TCRnine};
			
			\legend{AQP (1 user), SQP (1 user), AQP (10 users), SQP (10 users)}
		\end{axis}
	\end{tikzpicture}
	\caption{TCR9}
	\label{fig:ldbc:tcr9}
\end{subcaptionblock}
\\
\begin{subcaptionblock}{.45\textwidth}
	\centering
	\begin{tikzpicture}[scale=0.7]
		\begin{axis}[
			ybar,
			ymode=log,
			ymin=100,
			ymax=100000,
			enlarge x limits=0.2,
			bar width=7,
			xtick=data,
			xticklabels from table={\TCReleven}{CPUs},
			grid,
			legend style={
				legend columns=2,
				font=\footnotesize
			},
			legend cell align={left},
			legend image code/.code={
				\draw [#1] (-0.125cm,-0.125cm) rectangle (0.125cm,0.125cm);
			}
			]
			
			\addplot [pattern=north west lines] table [y=AQP_1user, x=X] {\TCReleven};
			\addplot [fill=lightgray] table [y=SQP_1user, x=X] {\TCReleven};
			
			\addplot [pattern=north east lines] table [y=AQP_10users, x=X] {\TCReleven};
			\addplot [fill=gray] table [y=SQP_10users, x=X] {\TCReleven};
			
			\legend{AQP (1 user), SQP (1 user), AQP (10 users), SQP (10 users)}
		\end{axis}
	\end{tikzpicture}
	\caption{TCR11}
	\label{fig:ldbc:tcr11}
\end{subcaptionblock}
\begin{subcaptionblock}{.45\textwidth}
	\centering
	\begin{tikzpicture}[scale=0.7]
		\begin{axis}[
			ybar,
			ymode=log,
			ymin=100,
			ymax=100000,
			enlarge x limits=0.2,
			bar width=7,
			xtick=data,
			xticklabels from table={\TCRtwelve}{CPUs},
			grid,
			legend style={
				legend columns=2,
				font=\footnotesize
			},
			legend cell align={left},
			legend image code/.code={
				\draw [#1] (-0.125cm,-0.125cm) rectangle (0.125cm,0.125cm);
			}
			]
			
			\addplot [pattern=north west lines] table [y=AQP_1user, x=X] {\TCRtwelve};
			\addplot [fill=lightgray] table [y=SQP_1user, x=X] {\TCRtwelve};
			
			\addplot [pattern=north east lines] table [y=AQP_10users, x=X] {\TCRtwelve};
			\addplot [fill=gray] table [y=SQP_10users, x=X] {\TCRtwelve};
			
			\legend{AQP (1 user), SQP (1 user), AQP (10 users), SQP (10 users)}
		\end{axis}
	\end{tikzpicture}
	\caption{TCR12}
	\label{fig:ldbc:tcr12}
\end{subcaptionblock}%
	\caption{Query times (P90) in ms for LDBC queries. The $y$-axis is in logarithmic scale.}
	\label{fig:ldbc:all}
\end{figure}

\begin{figure}
	\centering
	\pgfplotstableread{%
X  CPUs  SQP_1user  AQP_1user
1  8     4208.40    3590.20
2  16    3873.80    3623.10
3  32    3609.10    3348.00
4  64    3402.90    3145.70
}\TCRthree%
\begin{subcaptionblock}{.45\textwidth}
	\centering
	\begin{tikzpicture}[scale=0.7]
		\begin{axis}[
			ybar,
			ymode=log,
			ymin=100,
			ymax=100000,
			enlarge x limits=0.2,
			bar width=7,
			xtick=data,
			xticklabels from table={\TCRthree}{CPUs},
			grid,
			legend style={
				legend columns=2,
				font=\footnotesize
			},
			legend cell align={left},
			legend image code/.code={
				\draw [#1] (-0.125cm,-0.125cm) rectangle (0.125cm,0.125cm);
			}
			]
			
			\addplot [pattern=north west lines] table [y=AQP_1user, x=X] {\TCRthree};
			\addplot [fill=lightgray] table [y=SQP_1user, x=X] {\TCRthree};
			\legend{AQP (1 user), SQP (1 user)}
		\end{axis}
	\end{tikzpicture}
	\caption{TCR3}
	\label{fig:ldbc:tcr3}
\end{subcaptionblock}
\pgfplotstableread{%
X  CPUs  SQP_1user  AQP_1user
1  8     4463.70    5725.40
2  16    4572.90    6204.60
3  32    4328.60    5884.10
4  64    3477.90    4706.00
}\TCRfive%
\begin{subcaptionblock}{.45\textwidth}
	\centering
	\begin{tikzpicture}[scale=0.7]
		\begin{axis}[
			ybar,
			ymode=log,
			ymin=100,
			ymax=100000,
			enlarge x limits=0.2,
			bar width=7,
			xtick=data,
			xticklabels from table={\TCRfive}{CPUs},
			grid,
			legend style={
				legend columns=2,
				font=\footnotesize
			},
			legend cell align={left},
			legend image code/.code={
				\draw [#1] (-0.125cm,-0.125cm) rectangle (0.125cm,0.125cm);
			}
			]
			
			\addplot [pattern=north west lines] table [y=AQP_1user, x=X] {\TCRfive};
			\addplot [fill=lightgray] table [y=SQP_1user, x=X] {\TCRfive};
			\legend{AQP (1 user), SQP (1 user)}
		\end{axis}
	\end{tikzpicture}
	\caption{TCR5}
	\label{fig:ldbc:tcr5}
\end{subcaptionblock}%
	\caption{P90 processing times (in ms) for queries TCR3 and TCR5, for a single users. The $y$-axis is in logarithmic scale.}
	\label{fig:ldbc:tcr3-5}
\end{figure}

The results for the \textit{Finbench} benchmark are summarized by Fig.~\ref{fig:ldbc:all} and Fig.~\ref{fig:ldbc:tcr3-5}, while the raw results are reported in tabular format in Appendix~\ref{app:ldbc-benchmark-results}. Results show that Siren Federate processes complex graph queries in seconds, never exceeding the one-minute threshold, even with 10 concurrent users. In some cases, the system achieves sub-second response times (e.g., TCR1, TCR2, TCR11, and TCR12 with a single user). In the following, we first focus on queries TCR1, TCR2, TCR7, TCR9, TCR11, and TCR12 whose results are shown in Fig.~\ref{fig:ldbc:all}.

All queries except TCR11 require selecting fields from multiple entities, necessitating a sequence of one or more inner joins. While we plan to introduce broadcast-based inner-join strategies in the near future, Siren Federate currently offers only the partitioned hash-join strategy for processing inner joins. This choice is justified by the fact that distributed inner joins often require exchanging a significant amount of data, making a partitioned exchange approach more suitable than broadcast-based ones in most cases. Consequently, AQP and SPQ exhibit similar performance on most queries, as they generate identical query plans.

TCR11 selects fields from a single entity, allowing its execution to be performed using only semi-joins, for which Siren Federate offers multiple join strategies. For this query, AQP achieves better response times than SPQ -- 50\% faster on average in the single-user scenario -- as it takes into account statistics of intermediate join results to better select the appropriate join strategies. This optimization is particularly critical for TCR11, as it requires matching a graph pattern that can expand from a length of 1 up to 10, requiring a minimum of 4 joins and up to a maximum of 22 joins to be processed. When handling such a high number of joins, planning errors can easily accumulate, leading to significant latency. AQP mitigates this by leveraging finer-grained estimations, resulting in more efficient execution plans.

As the number of CPUs increases, we observe that query latency decreases for both AQP and SQP. However, in single‑user experiments, SQP is often slightly faster than AQP due to its higher parallelization. In fact, SQP determines strategies for all joins upfront, without considering intermediate. This enables greater operation-level parallelism across the physical execution. To better understand, consider the query plan in Fig.~\ref{fig:query-plan}. With SQP, the leftmost \textsf{SCAN A} in Stage 1 can execute concurrently with the \textsf{SCAN C} in Stage 3, even though Stage 3 depends on the results of Stage 1. This is possible because SQP pre-schedules all operations, allowing scans to pre-fetch data before than the joins which consume their results are ready to execute. In contrast, our current AQP implementation processes stages sequentially. For instance, AQP must complete the entire Stage 1 (including the join and all associated scans) before initiating any operations in Stage 3 when executing the query in Fig.~\ref{fig:query-plan}. The better parallelism provided by SQP can improve performance for single queries, though AQP more efficient join selection often compensates under concurrent workloads. In fact, the cost of the query plan remains similar (or better) with AQP. Consequently, under concurrent workload, the system can utilize underused resources across all running queries. In other words, while AQP may not saturate the CPU for a single query, under concurrent load it enjoys effectively the same (or better) throughput because the total workload is distributed more evenly. These findings suggest that further tuning of the adaptive planner -- particularly a mechanism to increase or decrease task parallelism based on the current load -- would improve the single-user performance of AQP without sacrificing its advantages under higher concurrency.

In the single-user run, results show a slight increase in latency for queries TCR1, TCR2, TCR7, TCR9, and TCR12 when moving from 32 to 64 CPUs. By contrast, this does not appear in the multi-user scenario. This suggests that once the workload saturates roughly 32 CPUs, adding more threads can actually incur extra overhead (e.g., for scheduling and coordinating join tasks) without delivering further speedup. Specifically, when the dataset is relatively small, distributing its partitions (morsels) across numerous threads yields diminishing returns -- each morsel becomes so small that parallelization overhead outweighs any gains. Under concurrent workloads, however, the higher core count is still well utilized because multiple queries can keep all cores busy.

Finally, we discuss the results for TCR3 and TCR5, which are shown in Fig.~\ref{fig:ldbc:tcr3-5}. Both queries require finding specific paths and are processed using the semi-join decomposition (SJD) technique from Sec.~\ref{sec:sjd}.
In the single-user run, Siren Federate returns results for both queries in a matter of seconds. Notably, this is thanks to SJD, as an implementation using the chains of inner-joins was unable to complete the queries. Such naive implementation was failing in over half of the runs due to the excessive memory consumption caused by the combinatorial growth of intermediate results. This highlights the benefits of SJD that can be explained by
\begin{inparaenum}[(1)]
	\item the presence of a reachability test to prevent unnecessary computation while expanding a path;
	\item the availability of different semi-joins strategies to handle different complexities in terms of data to be joined; and
	\item the fact that, at the time of writing, neither query planners support the ability to push \textsf{JOIN} operations to left \textsf{SCAN}s to reduce the amount of data to process.
\end{inparaenum}

As with the previous queries, we observe that AQP and SQP exhibit similar performance for TCR3, whereas SQP is significantly faster than AQP for TCR5. Examining the query plan execution logs reveals that SQP spawned a higher number of parallel tasks in certain phases, despite both planners having comparable per-task costs. This parallelization advantage impacts TCR5 because it executes joins that are more computationally intensive than the joins in TCR3. This mirrors our earlier observations that the lower task-parallelism of AQP can increase latency in single-user scenarios, whereas SQP tends to distribute the same workload across more threads, thereby completing faster when system resources are not contended by multiple concurrent workload.

Unfortunately, we cannot comment on the results for 10 concurrent users, as the system was unable to process them due to memory constraints. We believe this limitation is specific to the experimental setting and will be mitigated with additional main memory, either on a single computing instance or across multiple instances.

Our evaluation on the LDBC Finbench benchmark demonstrates the effectiveness of semi-join decomposition and adaptive query planning for complex graph queries. Building on these results, future work could explore: (1) testing with larger-scale Finbench datasets to assess SJD scaling properties across different graph structures; (2) comparative analysis with BFS-based traversal to better understand the performance trade-offs between approaches; and (3) exploring how semi-join decomposition could be applied to a broader range of graph queries, with the goal of developing a comprehensive cost model that would enable the query planner to automatically select the optimal strategy for each query type.

\subsection{Real-World Deployment Insights}
\label{sec:evaluation:real-world}

To validate our system's practical applicability, we examined, among others, a real-world deployment at Apollo.io\footnote{\url{https://siren.io/case-study-transforming-enterprise-search-at-apollo-with-siren-federate/}} involving a 350-node Elasticsearch cluster processing nearly half a petabyte of data with multiple concurrent users. The deployment demonstrated Siren Federate's ability to handle intricate data relationships and high-concurrency scenarios, demonstrating the effectiveness of our distributed join and query optimization techniques. Empirical results showed a dramatic reduction in query response time from an average of 7 seconds to sub-second, while significantly improving cluster stability and reducing query failures. These insights complement our quantitative benchmarks by providing empirical validation of Siren Federate's architectural design, specifically its approach to bridging document-oriented and relational data models in large-scale distributed environments.

\section{Conclusion}

This paper presented Siren Federate, a system that enables efficient exploratory analysis of large-scale knowledge graphs by bridging document-oriented, relational, and graph data models. Our architecture addresses  challenges faced by investigative intelligence workflows that require both advanced search capabilities and complex graph operations at scale.

Key contributions include integrating relational join operations within the document-oriented model, leveraging IR system capabilities, and implementing distributed join algorithms optimized for IR systems. We also introduced adaptive query planning for accurate runtime cardinality estimation, query plan folding to reduce redundant computations, and semantic caching to enhance iterative query performance. Columnar in-memory processing and Elasticsearch's log-structured distributed architecture were also highlighted.

An important contribution is a approach to path queries through Semi-Join Decomposition (SJD). SJD reduces the combinatorial explosion of intermediate results by breaking down path queries into semi-joins. This approach minimizes memory usage and computational overhead, making it more scalable and efficient for large graphs. While SJD is applicable to general multi-hop path queries, it is particularly effective in the context of all-shortest-paths problems, offering a practical solution to the challenges faced by traditional methods. SJD's integration with Siren Federate's adaptive query planner and semantic caching further enhances its effectiveness for exploratory graph analysis.

We validated the effectiveness of our architecture and techniques through two evaluation scenarios. First, using a synthetic dataset containing billions of documents, we demonstrated the system’s capability to sustain sub-second to second response times. Second, leveraging the LDBC Financial Benchmark, we assessed the system’s proficiency in efficiently handling complex graph queries in a matter of seconds. Additionally, we presented insights from a real-world deployment involving 350 nodes, highlighting the practical applicability and robustness of the system architecture under operational conditions.

In conclusion, our architecture retains the search and relevance ranking capabilities of the IR system while introducing efficient relational operations and graph analytics at scale. This demonstrates how a combination of document and relational system features can enhance scalability and analytical capabilities for exploratory analysis of large knowledge graphs. Looking forward, our research will focus on further enhancing parallel processing across data segments, extending the suite of available graph analytic techniques, and refining adaptive optimization methods to continually improve performance and usability. Additionally, we plan to further develop Semi-Join Decomposition by exploring broader applicability to complex graph patterns, and more rigorous theoretical analysis. Future work will also involve comparative benchmarking of SJD against alternative graph querying approaches to clearly delineate its strengths and boundaries of effectiveness.

\section*{\ackname}
We would like to thank Martin Anseaume, George Apaaboah, Issac Garcia, David Homes, Johnny Hujol, Flavio Pompermaier, and all past members of our engineering team for their significant contributions to the development of Siren Federate. Their efforts were essential to the realization of the system described in this work.

We used ChatGPT to improve the writing in some parts of this otherwise original manuscript. All AI-generated text has been carefully reviewed to ensure accuracy and correctness. 

\bibliographystyle{splncs03}
\bibliography{mybibliography}

\appendix
\section{LDBC Finbench}

This section reports the queries and benchmark results discussed in the evaluation Section~\ref{sec:evaluation:ldbc}.

\subsection{GQL Queries}\label{app:ldbc-gql-queries}

LDBC introduces several workloads for measuring different aspects of a graph engine. We express some of those workload as GQL queries that Siren Federate can interpret and internally translate into query plans for processing. We here lists the queries, for the complex translation workload, that we used in our experiments.

These queries are parameterized with a \textsf{START} and \textsf{END} variables to represent date ranges, and a \textsf{PERSON\_ID} and \textsf{ACCOUNT\_ID} variables to identify people and accounts. \textsf{WHERE} clauses -- used by GQL to filter entities that match a node pattern -- are here expressed using the Lucene query syntax. Finally, \textsf{SELECT} statements are amended to return raw data points since aggregation operators such as \textsf{SUM} are not supported yet at the time of writing.

\begin{lstlisting}[style=gql,language=SQL,caption=TCR1]
SELECT other.id, medium.mediumType, medium.id
FROM "ldbc-finbench"
MATCH (medium:Medium WHERE "isBlocked:true")
      ->(:MediumSignInAccount)
      ->(other:Account)
      (
        (:Account)
        <-(:AccountTransferAccount WHERE "createTime:{START TO END}")
        <-(:Account)
      ){1,3}
      (account:Account WHERE "id:ACCOUNT_ID")
\end{lstlisting}

\begin{lstlisting}[style=gql,language=SQL,caption=TCR2]
SELECT other.id, l.loanAmount, l.balance
FROM "ldbc-finbench"
MATCH (l:Loan)
      ->(:LoanDepositAccount WHERE "createTime:{START TO END}")
      ->(other:Account)
      (
        (:Account)
        ->(:AccountTransferAccount WHERE "createTime:{START TO END}")
        ->(:Account)
      ){1,3}
      (:Account)<-(:PersonOwnAccount)<-(person:Person WHERE "id:PERSON_ID")
\end{lstlisting}

\begin{lstlisting}[style=gql,language=SQL,caption=TCR3]
SELECT trace
FROM "ldbc-finbench"
MATCH trace = ALL SHORTEST (src:Account WHERE "id:ACCOUNT_ID")
      (
        (:Account)
        ->(:AccountTransferAccount WHERE "createTime:{START TO END}")
        ->(:Account)
      ){1,10}
      (dst:Account WHERE "id:ACCOUNT_ID")
\end{lstlisting}

\begin{lstlisting}[style=gql,language=SQL,caption=TCR5]
SELECT trace
FROM "ldbc-finbench"
MATCH trace =
      (
        (:Account)
        <-(:AccountTransferAccount WHERE "createTime:{START TO END}")
        <-(:Account)
      ) {1,3}
      (account:Account)<-(:PersonOwnAccount)<-(:Person WHERE "id:PERSON_ID")
\end{lstlisting}

\begin{lstlisting}[style=gql,language=SQL,caption=TCR7]
SELECT src.id, dst.id, edge1.amount, edge2.amount
FROM "ldbc-finbench"
MATCH (src:Account)
      ->(edge1:AccountTransferAccount WHERE "amount:{0 TO *} AND createTime:{START TO END}")
      ->(mid:Account WHERE "id:PERSON_ID")
      ->(edge2:AccountTransferAccount WHERE "amount:{0 TO *} AND createTime:{START TO END}")
      ->(dst:Account)
\end{lstlisting}

\begin{lstlisting}[style=gql,language=SQL,caption=TCR9]
SELECT edge1.amount, edge2.amount, edge3.amount, edge4.amount
FROM "ldbc-finbench"
MATCH (up:Account)
      ->(edge3:AccountTransferAccount WHERE "amount:{0 TO *} AND createTime:{START TO END}")
      ->(mid:Account WHERE "id:PERSON_ID")
      ->(edge4:AccountTransferAccount WHERE "amount:{0 TO *} AND createTime:{START TO END}")
      ->(down:Account)
      (mid)<-(edge1:LoanDepositAccount WHERE "amount:{0 TO *} AND createTime:{START TO END}")
      (mid)<-(edge2:AccountRepayLoan WHERE "amount:{0 TO *} AND createTime:{START TO END}")
\end{lstlisting}

\begin{lstlisting}[style=gql,language=SQL,caption=TCR11]
SELECT l.loanAmount
FROM "ldbc-finbench"
MATCH (l:Loan)<-(:PersonApplyLoan)<-(p2:Person)
      (
        (:Person)<-(g:PersonGuaranteePerson WHERE "createTime:{START TO END}")
        <-(:Person)
      ) {1,10}
      (p1:Person WHERE "id:PERSON_ID")
\end{lstlisting}

\begin{lstlisting}[style=gql,language=SQL,caption=TCR12]
SELECT compAcc.id, edge2.amount
FROM "ldbc-finbench"
MATCH (company:Company)
          ->(:CompanyOwnAccount)
          ->(compAcc:Account)
          <-(edge2:AccountTransferAccount WHERE "createTime:{START TO END}")
          <-(pAcc:Account)
          <-(:PersonOwnAccount)
          <-(person:Person WHERE "id:PERSON_ID")
\end{lstlisting}

\subsection{Raw Benchmark Results}
\label{app:ldbc-benchmark-results}

\begin{table}
	\centering
	\caption{Query times (P90) in ms for LDBC queries.}
\begin{tabular*}{\textwidth}{@{\extracolsep{\fill}} lllrrrrrrrr}
    \toprule
    Planner                  & Users               & CPUs & TCR1      & TCR2     & TCR3                  & TCR5                 & TCR7     & TCR9     & TCR11    & TCR12     \\
    \cmidrule{2-11}
    \multirow{8}{*}{Adaptive}& \multirow{4}{*}{1}  & 8    & 1,428.8   & 818.3    & 3,590.2               & 5,725.4              & 6,775.2  & 5,263.5  & 200.1    & 244.9   \\
                             &                     & 16   & 666.7     & 732.0    & 3,623.1               & 6,204.6              & 6,521.0  & 4,983.0  & 140.0    & 223.0   \\
                             &                     & 32   & 685.0     & 727.9    & 3,348.0               & 5,884.1              & 5,732.5  & 4,171.3  & 132.9    & 221.0   \\
                             &                     & 64   & 867.6     & 909.8    & 3,145.7               & 4,706.0              & 6,756.6  & 4,742.8  & 133.0    & 298.0   \\
    \cmidrule{2-11}                                                                                                                                   
                             & \multirow{4}{*}{10} & 8    & 25,453.7  & 22,193.7 & \multirow{4}{*}{N/A}  & \multirow{4}{*}{N/A} & 25,838.8 & 14,864.2 & 12,905.5 & 8,276.4 \\
                             &                     & 16   & 9,287.7   & 8,676.2  &                       &                      & 17,073.9 & 11,452.4 & 5,050.8  & 3,074.9 \\
                             &                     & 32   & 1,655.9   & 1,760.9  &                       &                      & 10,057.9 & 7,046.9  & 911.7    & 667.0   \\
                             &                     & 64   & 1,256.0   & 1,397.0  &                       &                      & 9,061.8  & 6,374.9  & 280.9    & 426.0   \\
    \midrule                                                                                                                                          
    \multirow{8}{*}{Static}  & \multirow{4}{*}{1}  & 8    & 1,779.5   & 1,199.0  & 4,208.4               & 4,463.7              & 7,551.8  & 5,690.6  & 337.0    & 347.9   \\
                             &                     & 16   & 945.8     & 798.9    & 3,873.8               & 4,572.9              & 7,344.9  & 5,244.0  & 325.9    & 241.9   \\
                             &                     & 32   & 611.0     & 594.7    & 3,609.1               & 4,328.6              & 5,271.0  & 3,627.9  & 288.9    & 179.0   \\
                             &                     & 64   & 835.9     & 644.8    & 3,402.9               & 3,477.9              & 6,184.2  & 4,181.5  & 287.5    & 209.9   \\
    \cmidrule{2-11}                                                                                                                                   
                             & \multirow{4}{*}{10} & 8    & 26,001.4  & 24,126.6 & \multirow{4}{*}{N/A}  & \multirow{4}{*}{N/A} & 28,029.8 & 15,520.8 & 14,210.3 & 8,943.4 \\
                             &                     & 16   & 9,994.3   & 10,071.6 &                       &                      & 16,677.9 & 11,782.8 & 5,837.9  & 3,554.8 \\
                             &                     & 32   & 2,311.8   & 1,635.0  &                       &                      & 9,837.9  & 7,270.9  & 1,661.5  & 1,289.6 \\
                             &                     & 64   & 1,235.9   & 1,131.9  &                       &                      & 8,712.8  & 5,941.0  & 499.0    & 369.0   \\
    \bottomrule
\end{tabular*}
	\label{tab:results_ldbc}
\end{table}

\end{document}